\documentclass[reprint,amsmath,amssymb,aps,]{revtex4-1}

\usepackage{graphicx}% Include figure files
\usepackage{dcolumn}% Align table columns on decimal point
\usepackage{bm}% bold math
\usepackage{braket}
\usepackage{xcolor}
\usepackage{footmisc}
\usepackage{amsmath}

\begin{document}

\title{First-principle evolution Hamiltonian operator: derivation from ADM quantum constraints and quantum reference-frame conditions}

\author{Chun-Yen Lin}
\email{10050031@gm.scu.edu.tw}
\affiliation{Department of Physics, Soochow University, Taipei 111, Taiwan}

\begin{abstract}
For any Dirac theory of quantum gravity governed by a set of well-defined quantum constraints, we discover a universal formula for the exact form of the evolution Hamiltonian operator in a variable quantum reference frame of our construction, expressed in terms of the quantum-constraint operators and frame-condition operators as the only inputs. Due to the first-principle nature of the formula, the evolution Hamiltonian operator contains the full interactions encoded in the quantum constraints, and it generates the Schr\"odinger evolution described by the genuine quantum relational observables associated to the frame and acting in the physical Hilbert space solving the quantum constraints.
\end{abstract}

\maketitle

%\tableofcontents

\section{Introduction}

For the canonical mechanics in a fixed background spacetime, the Hamiltonian generates the consistent translations of all the interacting objects along the time direction of the given spacetime, thereby giving the evolution of the objects in the given notion of time. For the canonical formulation of general relativity (canonical GR)\cite{1.1}, the interacting objects include the spacetime itself appearing as the gravitation fields. Accordingly, the Hamiltonian generates the consistent translations of the whole universe--including the spacetime itself-- in the time-coordinate direction of an abstract spacetime coordinate system. Thereby, the Hamiltonian in canonical GR generates instead the coordinate transformations, which are the local invariance symmetries instead of the physical evolutions. Indeed, as the local symmetry generator, the Hamiltonian appears as a linear combination of the set of first-class constraints $\{C_\mu=0\}$ known as the Arnowitt–Deser–Misner constraints (ADM constraints)\cite{1.2,1.1}, over the arbitrary coefficients of Lagrange multipliers. 

In the most prevailing point of views \cite{2.1,3.1,4}, the quantum theory of canonical GR is then governed by the quantized ADM constraints $\{\widehat{C}_\mu\}$, which generate the fundamental symmetries at the quantum level and define the physical quantum states by annihilating them. Under the various formulations of quantum canonical GR following such views, different challenges \cite{19.1} arise for obtaining the unitary evolution along a valid notion of time. 

The Wheeler-DeWitt (WDW) formulation\cite{2.1,2.2 pathint wdweq,5 wdweq schrdgr} is based on the study of the quantum constraint equations as the wave function equations of the universe, called the WDW equations. It has been long observed that these equations take the “interacting Klien-Gordon form”, instead of the Schr\"odinger equation for the physical evolution. Progresses\cite{2.1,5 wdweq schrdgr} have been made in relating the two crucial equations, and it is shown that by appplying the Wentzel–Kramers–Brillouin (WKB) or Born-Oppenheimer (BO) approximations to proper wavefunctions solving the WDW equation, one may perturbatively factor out the (approximate) Schr\"odinger evolution of a ``light perturbative sector” against the background time provided by the ``heavy sector”. In the canonical path-integral formulation\cite{3.2path,2.2 pathint wdweq,6 pathint superprop canon wkb} based on the Einstein-Hilbert action, the integral has been studied as the so-called superspace propagator implicitly solving the quantum constraints $\{C_\mu\}$; it is known that due to the dynamic nature of the gravitation fields in the action, the superspace propargator is non-unitary and fundamentally distinct from the Schr\"odinger propagator. Again, under the BO or WKB approximations, it is shown\cite{3.2path,6 pathint superprop canon wkb} that the approximate Schr\"odinger unitary propagator for the perturbative sector can be factorized from the superspace propargator, and the unitary evolution against the semiclassical background of the heavy sector can be obtained perturbatively. 

In order to explore the strong gravitational interactions in deep quantum regimes with high-precision, the main challenge for the above two broad formulations is toward a general non-perturbative method beyond the the BO or WKB approximations. Importantly, newer insights\cite{7.1,7.2} have been brought to the precise relation of the Schr\"odinger propagator in a certain ``clock time” to the Einstein-Hilbert path integral in the corresponding ``clock-gauge”; these results strongly suggest a fundamental translation between the ``properly gauge-fixed” superspace propagators and the associated Schr\"odinger propagators referring to the same clocks.

In the modern developments\cite{4,8.1 raq,8.2 lqg,15.2} of the so-called Dirac formulation, the fundamental quantum constraint operators $\{\widehat{C}_\mu\}$ are explicitly constructed for the non-perurbative physical state space to be defined as the common kernel of the constraints. The kernel has been rigorously defined as the image of the constraint-kernel projector known as the constraint rigging map\cite{8.1 raq,8.2 lqg,9.1 rigg}, whose matrix elements also provide the inner product for the kernel as a Hilbert space. Here the challenge lies in the concrete characterization of the physical Hilbert space\cite{8.2 lqg} and finding the spacetime-localized quantum observables\cite{8.3 obsv,8.4 obsv,8.5 obsv}that capture the unitary evolution, which has to be computable without giving up the full interactions the Dirac formulation is introduced for.

Remarkably, the modern Dirac formulation seems to be the more fundamental and play a unifying role amongst the other approaches. Indeed, it is known\cite{10.1 pathint rigg} that the rigging map matrix elements can be written as a gauge-fixed Einstein-Hilbert path integral with the corresponding boundary states. Also, the physical Hilbert space as the image of the rigging map is annihilated by the quantum constraints, so it is the non-perturbative wave-function solution space to the associated WDW equations. Thus, all of the challenges mentioned above may be tackled in the framework of the modern Dirac formulation, in which we may search for the first-principle method of deriving the unitary evolutions directly from the quantum constraints.

Under such motivations, our approach \cite{11,12,13} of quantum reference frames aims for constructing the ``relational quamtum observables"\cite{8.5 obsv} in the Dirac formulation, via deep quantum notion of reference frames utilizing the ``reference fields" from a chosen ``reference sector" of the unconstrained kinematic Hilbert space. Working within the Dirac formulation via the mentioned rigging map method, the construction assumes that the quantum constraints are implemented by the constraint rigging map $\widehat{\mathbb{P}}\equiv \delta(\,\sum_\mu \widehat{C}^2_\mu\,)$, projecting the kinematic Hilbert space into the physical Hilbert space, where the physical inner product is defined by the associated matrix elements of $\widehat{\mathbb{P}}$. Based on the kinematic Hilbert space with the elementary set of operators $\{\widehat{X}_\mu, \widehat{P}_\mu, \widehat{X}_I, \widehat{P}_I\}$, a quantum reference frame is specified with the instantaneous eigenspaces $t\mapsto \mathbb{K}_t$ of the reference fields $\widehat{T}_\mu$ constructed with the reference sector’s elementary operators $\{\widehat{X}_\mu, \widehat{P}_\mu\}$.  The reference field eigenspace $\mathbb{K}_t$ of each moment $t$ is required to be isomorphic to its image in the physical Hilbert space under the rigging map. The isomorphisms then naturally push foward the elementary set $\{\widehat{X}_I, \widehat{P}_I\}$ for the $\mathbb{K}_t$, into the elementary relational operators $\{\widehat{X}_I(t), \widehat{P}_I(t)\}$ as the operators in the physical Hilbert space\cite{13}, detecting the values the of ${X}_I$ and ${P}_I$ at the moment of the universe when the reference fields take the values of ${T}_\mu(t)$ without uncertainty. As to be discussed, in a given quantum reference frame the reference fields effectively act as a set of non-dynamical parameters appearing in the relational-evolution laws for the dynamical sector, distinguishing our approach from those with the reference fields as the ``quantum clocks" with their own quamtum dynamics.

Further, for the elementary relational operators as Heisenberg operators the associate propagator $\widehat{U}_{t_2 t_1}$, as a square matrix, is composite of the square matrices $\widehat{P}_{t_2 t_1}$ given by the matrix elements of $\widehat{\mathbb{P}}$ between the $\mathbb{K}_{t_1}$ and $\mathbb{K}_{t_2}$. Explicitly, the obtained result \cite{12,13} is $\widehat{U}_{t_2 t_1}= \widehat{P}^{-1/2}_{t_2t_2} {P}_{t_2 t_1} \widehat{P}^{-1/2}_{t_1 t_1}$. In view of the path integral formulation, the middle matrix $\widehat{P}_{t_2 t_1}$ is the Hilbert-Einstein path intergral in the gauge with the specified $t\mapsto{T}_\mu(t)$. It is showned \cite{13} that the sandwitching matrices $\widehat{P}^{-1/2}_{t_2 t_2}$ and $\widehat{P}^{-1/2}_{t_1 t_1}$ serve as the operator realization of the mentioned transformation from the path-integral into the Schr\"odinger propagator proposed in \cite{7.1}. Since the $\widehat{U}_{t_2 t_1}$ defines the unitary transformation between $\{\widehat{X}_I(t_2), \widehat{P}_I(t_2)\}$ and $\{\widehat{X}_I(t_1), \widehat{P}_I(t_1)\}$, it is generated by a corresponding time-dependent evolution Hamiltonian $\widehat{H}_t$ operator built from the set $\{\widehat{X}_I, \widehat{P}_I\}$. With $\widehat{P}_{t’ t}$ as the basic building blocks, the form of $\widehat{H}_t$ must be  determined by the quantum constraints $\widehat{C}_\mu$ and the reference frame conditions $t\mapsto \mathbb{K}_t$. In this first principle manner, the operator $\widehat{H}_t$ containing the full interactions is defined without relying on any semiclassical or perturbative approximations. 

In this work, we accomplish the next crucial step to show that the $\widehat{H}_t$ is not only fundamentally defined at the abstract level, but also readily computable with a universal formula. To do so, we introduce the Wigner-Weyl representation \cite{14 wigner} to our construction, so that each kinematic operators $\widehat{A}=\widehat{A}( \widehat{X}_\mu, \widehat{P}_\mu, \widehat{X}_I, \widehat{P}_I )$ is uniquely represented by a phase-space function $G_{\scriptscriptstyle{\widehat{A}}}( X_\mu, P_\mu, X_I, P_I )$, while the operator algebra is represented by the non-commutative ``$\star$-product" between the phase space functions. Under this representation for our construction, via the functions $G_{\scriptscriptstyle{\widehat{P}_{t’ t}}}(X_I, P_I )$ as the basic building blocks, we derive the evolution Hamiltonian $G_{\scriptscriptstyle{\widehat{H}_t}}( {X}_I, {P}_I )$, expressed in terms of the quantum constraints $G_{\scriptscriptstyle{\widehat{C}_\mu}}( X_\mu, P_\mu, X_I, P_I )$ the and the frame conditions consisting of $ G_{\scriptscriptstyle{\ket{ X_\mu(t)}\bra{ X_\mu(t)}}} (X_\mu, P_\mu)$ and $G_{\scriptscriptstyle{\widehat{f}}} (X_\mu, P_\mu)$. From a given initial data for the elementary relational operators $G_{\scriptscriptstyle{\widehat{X}_I(0)}}$ and $G_{\scriptscriptstyle{\widehat{P}_I(0)}}$ the Hamiltonian $G_{\scriptscriptstyle{\widehat{H}_t}}$ can then be used to generate-- via the $\star$-bracket as the deformed Poisson bracket-- the evolution flow with the full quantum gravitational interactions encoded in the governing quantum constraints. Also, it will be demonstrated that our construction leads to a power series expansion of the $G_{H_t}(X_I, P_I )$, with the leading order term successfully reproducing the classical evolution Hamiltonian under the corresponding reference frame in the canonical GR. 

To the author's knowledge, this work gives the first explicit derivation of the evolution Hamiltonian operator with such signifcance and meanings.

\section{Quantum reference frame for Dirac theory }\label{s2}

We first give an updated prescription of our quantum reference frame approach \cite{11,12,13}. All the unspecified letter indices should be treated as abstract indices, so that $X_i$ denotes the array of $(X_1,X_2,...)$; also, the abstract indices labeled with the same letter should be numerically matched when evaluating an equation.

We assume a Dirac theory of canonical quantum gravity based on either the ADM formulation of general relativity possibly coupled to matters, or its symmetry reduced cosmological models. The theory is purely governed by a set of quantum constraints $\{\widehat{C}_\mu\}$ containing the quantized ADM constraints, constructed as self-adjoint operators in the unconstrained kinematic Hilbert space $\mathbb{K}$. Existing examples from the full theories include the ones in geometrodynamics \cite{15.1,15.2} and loop quantum gravity \cite{15.3}, where the constraint system may also contain the Yang-Mills quantum constraints \cite{15.3} due to the internal gauge symmetries. We assume every linear maps in $\mathbb{K}$ can be algebraically constructed with only a set of elementary operators $\{\widehat{X}_i, \widehat{P}_i\}$ well-defined in a dense domain $\mathbb{S}\subset\mathbb{K}$. We also assume that the elementary operators can be devided into two complete sets conjugate to each other, so that they provide the associated  (eigen- or coherent-) bases with $\mathbb{K}=span\{\ket{X_i}\}=span\{\ket{P_i}\}$ in a proper sense that 
\begin{eqnarray}
\label{kin0}
\ket{\psi}= \int D{X}_i\, \psi({X}_i)\, \ket{X_i}=\int D{P}_i \,\psi({P}_i) \,\ket{P_i}
\end{eqnarray}
for any $\psi \in \mathbb{K}$, with their given spectra and measures $D{X}_i $ and $D{P}_i $ for the integrals.

Specifically, in the canonical cases of $\mathbb{K}$ obtained with the self adjoint operators $\{\widehat{X}_i\}$ and $\{\widehat{P}_i\}$ providing the canonical conjugate pairs labeled by the index $i$ for the field species and modes, the only non-vanishing elementary commutators are of the familiar form
\begin{eqnarray}
\label{kin1}
[\widehat{X}_j, \widehat{P}_k]=i \hbar  \,\delta_{jk}.
\end{eqnarray}
Here, the expansion \eqref{kin0} is given with the two corresponding eigenbases, over the common eigen spectrum of $\mathbb{R}$ and with the standard measure of $D{X}_i\equiv d{X}_i$ and $D{P}_i\equiv d{P}_i$. 

Our general setting also include the non-canonical cases of the $\mathbb{K}$ constructed from the so-called flux-holonomy quantization \cite{15.3}\cite{16 lqc} upon a single oriented graph. The given oriented graph-- consisting of a set of vertices linked by a set of directional edges-- is dual to a cell decomposition of the spatial manifold which is effectively discretized. Similar to the setting of lattice gauge theory, the $\{\widehat{X}_i\}\equiv \{\widehat{X}_{n,e}\}$ represent the holonomy variables over the edge $e$, defined with the connection fields (or the gauge fields) of the $n$th species and taking values in the associated compact Lie group $\mathcal{G}_n$. The conjugate variables $\{\widehat{P}_i\}\equiv \{\widehat{P}_{n,e,a}\}$ represent the flux variables over the cell face dual to the edge $e$, defined involving the triad fields (or the electric fields) of the $n$th species and taking values in the $a$th generator-component of the associated Lie algebra  $\mathfrak{g}_n$. With the holonomy operators expressed a chosen representation and the corresponding generator basis of $\mathfrak{g}_n$ given by $\{\tau^a\}$, the elementary flux-holonomy operators have the only non-vanishing commutators of 
\begin{eqnarray}
\label{kin2}
&&[\widehat{X}_{n,e}\,,\, \widehat{P}_{n,e',a}]=(i q_n \hbar  \,\delta_{e,e'})\, \tau^a\widehat{X}_{n,e}
\nonumber\\
&&[\widehat{P}_{n,e,a}\,,\,\widehat{P}_{n,e',b}]=(i q_n \hbar\,\delta_{e,e'})  \,\sigma_{abc}\,\widehat{P}_{n,e',c}
\end{eqnarray}
Here $q_n$ is the ``coupling constant " for the species, and $\sigma_{abc}$ is the structure constant of $\mathcal{G}_n$. In this case, the basis expansion \cite{16 lqc cohere} will take the form of 
\begin{eqnarray}
\label{kin0}
\ket{\psi}= \int D{X}_{n,e}\,\, \psi({X}_{n,e})\, \ket{{X}_{n,e}}=\int D\vec{P}_{n,e}\, \,\psi(\vec{P}_{n,e}) \,\ket{\vec{P}_{n,e}}.\nonumber\\
\end{eqnarray}
Here, the first expansion is using the holonomy eigenbasis, with the integration for each $(n,e)$ over the spectrum of $\mathcal{G}_n$ with the $D{X}_{n,e}$ being the Haar measure. The flux basis expansion is instead given by the resolution of identity using the spin-coherent states $\ket{\vec{P}_{n,e}}$ maximally peaked for $\{\widehat{P}_{n,e,a}\}$, over the peaked-value domain of $\mathfrak{g}_n$ with the invariant measure $D\vec{P}_{n,e}$ induced from the Haar measure of $\mathcal{G}_n$. 

We assume that all the self-adjoint quantum constraints  $\{\widehat{C}_\mu\}: \mathbb{S}\to \mathbb{S}$ are well-defined in the dense domain $\mathbb{S}\subset\mathbb{K}$. To impose the system of quantum constraints with one single kernel projector, we introduce the master constraint operator \cite{15.3} defined as
\begin{eqnarray}
\label{master}
\!\widehat{M}\equiv \sqrt{\sum_\mu \widehat{C}_\mu^2}: \mathbb{S}\to \mathbb{S}
\end{eqnarray}
so that the associated rigging map \cite{8.1 raq,9.1 rigg} operator $ \widehat{\mathbb{P}}$ is given by
\begin{eqnarray}
\label{rig}
 \widehat{\mathbb{P}}\equiv \delta(\widehat{M}) \,:\,\mathbb{S} \to \mathbb{S}^*\,\,;\,\,\ket{\psi} \mapsto \bra{\psi'}\equiv \bra{\psi} \widehat{\mathbb{P}}
\end{eqnarray}
where $\mathbb{S}^*$ denotes the algebraic dual of $\mathbb{S}$, and the $\bra{\psi} \widehat{\mathbb{P}}$ is the notation for the result of $\widehat{\mathbb{P}}$ acting on the $\ket{\psi}$. As a generalized kernel projector for the quantum constraint $\widehat{M}$, the rigging map serves two roles at once: the completion of its image $\bar{Img}(\widehat{\mathbb{P}})\equiv \mathbb{H}\subset \mathbb{S}^*$ gives the physical Hilbert space, with the Hermitian inner product for any two physical states $|\Psi_1)\equiv \bra{\psi_1}\widehat{\mathbb{P}}$ and $|\Psi_2) \equiv \bra{\psi_2}\widehat{\mathbb{P}}$ defined by
\begin{eqnarray}
\label{inner product}
(\Psi_1|\Psi_2)\equiv \left\langle {{\psi _1}} \right|\mathbb{P}\left| {{\psi _2}} \right\rangle.
\end{eqnarray}
Here, the value of $\bra{\psi_1}\widehat{\mathbb{P}}\ket{\psi_2}$ is given by the natural pairing between the $\bra{\psi_1}\widehat{\mathbb{P}}\in\mathbb{S}^*$ and $\ket{\psi_2}\in \mathbb{S}$. 

Note that the physical dynamics has to emerge from the physical states of ``quantum spacetimes" in $\mathbb{H}$ constructed without an absolute notion of time. Certainly, the key objects for the dynamics would be the physical inner product \eqref{inner product} which may be thought of as the `` kinematic transition amplitudes" between $\ket{\psi _1}$ and $\ket{\psi _2}$. Let us denote the set of subspaces $\{\mathbb{K}_n\subset \mathbb{K}\}$ with a dense domain $\mathbb{S}_n\subset\mathbb{K}_n$ satisfying $\mathbb{S}_n\subset\mathbb{S}$. For each pair $\mathbb{K}_1$ and $\mathbb{K}_2$ of the subspaces we may introduce a map between the associated $\mathbb{S}_1$ and $\mathbb{S}^*_2$ via the amplitudes as
\begin{eqnarray}
\label{transit}
\widehat{\mathbb{P}}_{21}: \mathbb{S}_1\to \mathbb{S}_2^*\,;\,\, \ket{\psi_1}\mapsto \bra{\psi_2}\nonumber\\
\braket{\psi_2| \phi_2}\equiv \bra{\psi_{1}}\widehat{\mathbb{P}}\ket{\phi_2}\,\,\forall \ket{\phi_2}\in\mathbb{S}_2 .
\end{eqnarray}
In general, the image of our rigging map $\widehat{\mathbb{P}}$ is not in $\mathbb{K}$ for important physical reasons. Indeed, each element in the image is a wavefunction in the common kernel for all the quantum ADM constraints and thereby invariant under the transformations generated by the quantum constraints. In the classical GR, the ADM constraints generates the orbits of coordinate transformations deforming one Cauchy surface into another. In the quantum theory, a wavefunction in the image of  $\widehat{\mathbb{P}}$ is constant valued along these potentially non-compact transformation orbits, and it would thus tend to be kinematically non-normalizable.  In another view,  the orbits manifest as the ``gauge degeneracy" in the map $\widehat{\mathbb{P}}$, which maps all the ``gauge-equivalent" elements in the same orbit into the same physical state in $\mathbb{S}^*$. Therefore, if the subspaces $\mathbb{K}_1$ and $\mathbb{K}_2$ is such that $\widehat{\mathbb{P}}_{11}$ and $\widehat{\mathbb{P}}_{22}$ becomes injective without any gauge degeneracy, we may expect the image of $\widehat{\mathbb{P}}_{21}$ to consist of only the kinematically normalizable wavefunctions . For the general set of domains $\{\mathbb{S}_n\}$ introduced earlier, we thus make a further assumption that 
 \begin{eqnarray}
\label{cont2}
\widehat{\mathbb{P}}_{11}, \,\widehat{\mathbb{P}}_{22}\,\, \text{injective}\,\,\,\, \Rightarrow\,\,\,\,\,,\, \,\mathbb{S}_{1}\xrightarrow[]{\widehat{\mathbb{P}}_{21}} \mathbb{S}_{2}\subset\mathbb{S}_{2}^* 
\end{eqnarray}
where we identify $\mathbb{S}_2$ with the subset of the continuous dual space $\mathbb{K}_2^{*cont}\subset \mathbb{K}_2^{*} \subset\mathbb{S}_2^*$ through the standard Reiz representation. This assumption has the following implications. By its given Hermicicity, the map $\widehat{\mathbb{P}}_{11}$ satisfying the above becomes a positive-definite and self-adjoint map in $\mathbb{K}_1$ with trivial-measured zeros in its spectral decomposition over the non-negative spectrum; with the $\mathbb{S}_1$ set to be the wave functions exponentially supressed toward the infinities and zeros in the decomposition, the unique inverse of $\widehat{\mathbb{P}}_{11}$ and its unique square root are given as $$\widehat{\mathbb{P}}^{-1}_{11}: \mathbb{S}_1\to \mathbb{S}_1\,\,\,\,\,\,\,\,\widehat{\mathbb{P}}^{-1/2}_{11}: \mathbb{S}_1\to \mathbb{S}_1$$ satisfying 
 \begin{eqnarray}
\label{diag}
\,\widehat{\mathbb{P}}^{-1/2}_{11}\,\widehat{\mathbb{P}}_{11}\,\widehat{\mathbb{P}}^{-1/2}_{11}=\widehat{I}_{11}\,: \mathbb{S}_1\to \mathbb{S}_1
\end{eqnarray}
with $\widehat{I}_{11}$ denoting the identity map. Under the assumed condition \eqref{cont2} for the rigging map, we introduce our approach of quantum reference frame through the algorithm consisting of the following steps.

\subsection{Algorithm under general setting}

To derive the physical dynamics, we now apply our quantum reference frame algorithm to the class of Dirac theories specified above. The steps are given by the following.

\begin{enumerate}
\item  To specify a quantum reference frame, we first choose from $\mathbb K$ a reference sector by splitting its elementary set into $$\{(X_i,P_i) \}= \{(X_\mu, P_\mu)\}\cup \{(X_I, P_I)\}, $$ with the two commuting sets $\{(X_\mu, P_\mu)\}$ and $\{(X_I, P_I)\}$ being respectively the elementary sets of the chosen ``reference sector" and ``dynamic sector". Then, with a dimensionless $t\in \mathbb{R}$ serving purely as a sequence parameter, a quantum reference frame is fully specified by a map $$t \mapsto \mathbb{K}_t\equiv span\{\ket{T_\mu(t),P_I}\}\subset \mathbb{K}$$ where $\mathbb{K}_t \in\{\mathbb{K}_n\}$ for each $t$ is a common eigenspace of a properly chosen set of reference fields $\{\widehat{T}_\mu=\widehat{T}_\mu(\widehat{X}_\nu, \widehat{P}_\nu)\}$ constructed with the reference sector’s elementary operators. Under such a map, we also introduce the summed spaces defined by $$\mathbb{K}_{\scriptscriptstyle{[t_1,t_2]}}\equiv span\big\{\ket{T_\mu(t),P_I}\big\}_{t\in[t_1,t_2]}\in \{\mathbb{K}_n\}$$. Setting  $\mathbb{S}_t\subset \mathbb{K}_t$ and  $\mathbb{S}_{\scriptscriptstyle{[t_1,t_2]}} \subset\mathbb{K}_{\scriptscriptstyle{[t_1,t_2]}}$ as the associated dense domains from $\{\mathbb{S}_n\}$, we denote their rigging-map images to be respectively $\mathbb{D}_t\subset\mathbb{H}$ and $\mathbb{D}_{\scriptscriptstyle{[t_1,t_2]}}\subset\mathbb{H}$, with the associated completions as the sub Hilbert spaces $\mathbb{H}_t\equiv \bar{\mathbb{D}}_t$ and $\mathbb{H}_{\scriptscriptstyle{[t_1,t_2]}}\equiv \bar{\mathbb{D}}_{\scriptscriptstyle{[t_1,t_2]}}$ .

\item The above map $t\mapsto\mathbb{K}_t$ is valid in defining a quantum reference frame for the duration $ [t_1,t_2] $, when the relevant transition maps ($t,t’\in [t_1,t_2] $)  
\begin{eqnarray}
\label{00}
\widehat{\mathbb{P}}_{t't}: \mathbb{S}_t\to \mathbb{S}^*_{t'}\,\,\,\text{and}\,\,\,\widehat{\mathbb{P}}_{t,\scriptscriptstyle{[t_1,t_2]}}: \mathbb{S}_{\scriptscriptstyle{[t_1,t_2]}}\to\mathbb{S}^*_{t}
\end{eqnarray}
defined as in \eqref{transit} satisfy the two crucial conditions
\begin{eqnarray}
\label{uni cond}
\mathbb{S}_t\xrightarrow[\text{injective}]{\widehat{\mathbb{P}}_{tt}} \mathbb{S}_t^* \,\,\Leftrightarrow\, \,\mathbb{S}_t \xrightarrow[\text{isomorphic}]{\widehat{\mathbb{P}}\restriction_{\mathbb{S}_{t}}} \mathbb{D}_t \\
\label{uni cond2}
\mathbb{S}_{\scriptscriptstyle{[t_1,t_2]}}\xrightarrow[\text{injective}]{\widehat{\mathbb{P}}_{t_1,\scriptscriptstyle{[t_1,t_2]}}} \mathbb{S}_{t_1}^* \,\,\Leftrightarrow \,\,\mathbb{H}_{t}= \mathbb{H}_{t_1}
\end{eqnarray}
Referring to the inner product for $\mathbb{H}$ given by \eqref{inner product}, it becomes clear that the above first condition is equivalent to that $\mathbb{P}\restriction_{\mathbb{S}_{t}}: \mathbb{S}_{t}\to \mathbb{D}_{t}$ is isomorphic for all $t$, while the second condition is equivalent to the triviality of the orthogonal complement of $\mathbb{D}_{t_1}$ as a subspace of $\mathbb{D}_{\scriptscriptstyle{[t_1,t_2]}}$-- which means $\mathbb{D}_{t_1}$ is dense in $\mathbb{D}_{\scriptscriptstyle{[t_1,t_2]}}$ and thus the two sets have the same completion $\mathbb{H}_{t}= \mathbb{H}_{t_1}$. Hence, we will call these two conditions the isomorphism and domain stability conditions.

\item Suppose that a valid quantum reference frame is chosen. We then have \eqref{diag} with $\widehat{\mathbb{P}}_{11}\equiv\widehat{\mathbb{P}}_{tt}$, which with the inner product provided in \eqref{inner product} implies that
\begin{eqnarray}
\label{isom}
\mathbb{P}\,\mathbb{P}^{-1/2}_{tt}\equiv \widehat{I}_t \restriction_{{S}_{t}} \,:\,\,\mathbb{S}_{t} \xrightarrow[]{\text{isometry}} \mathbb{D}_{t}\,\nonumber\\
\widehat{I}_t \,:\,\,\mathbb{K}_{t} \xrightarrow[]{\text{isometry}} \mathbb{H}_{t}\,,
\end{eqnarray}
where the map $\widehat{I}_t$ defines an isometry map between $\mathbb{K}_{t}$ and $\mathbb{H}_{t}$ as the unique contiuous extension of the $\widehat{I}_t \restriction_{{S}_{t}}$. Through this map, any operator $\widehat{\mathcal{X}}(\widehat{X}_I,\widehat{P}_I):\mathbb{S}_{t}\to\mathbb{S}_{t}$ is directly push forward into its isometric copunterpart in $\mathbb{D}_{t}$ as 
\begin{eqnarray}
\label{obsv0}
\widehat{\mathcal{X}}(t) \equiv  \widehat{I}_t \,\widehat{\mathcal{X}} \,\widehat{I}^{-1}_t\,\,:\,\,\mathbb{D}_{t}\to\mathbb{D}_{t}
\end{eqnarray}
densely defined in $\mathbb{H}_{t}$. We will call the set $\{(X_I(t_1), P_I(t_1))\}$ the elementary relational operators at the moment $t_1$. By the domain stability condition, the sets $\{(X_I(t), P_I(t))\}$ for all $t\in [t_1,t_2] $ are all densely defined in the same Hilbert space of $\mathbb{H}_{t}=\mathbb{H}_{t_1}$, serving as a Dirac state space describable by the elementary relational operators at the various moments. 

\end{enumerate}

The construction has several important features and corresponding open questions on its physical meanings, which we now describe in below.

\subsubsection*{ Quantum degrees of freedom in $\mathbb{D}_{t_1}$ and elementary relational observables}

Our construction shows that for any quantum reference frame $t\mapsto\mathbb{K}_t$ the degrees of freedom in the described Hilbert space $\mathbb{H}_{\scriptscriptstyle{t_1}}$ are indistinguishable from those of the $\mathbb{K}_t$. Indeed, by the isometry construction, the operator composition is clearly preserved in the sense that 
\begin{eqnarray}
\label{comp}
\widehat{\mathcal{A}}(\widehat{{P}}_J(t),\widehat{{X}}_I(t) )=[\widehat{\mathcal{A}}(\widehat{{P}}_J,\widehat{{X}}_I )](t) : \mathbb{D}_t\to \mathbb{D}_t
\end{eqnarray}
where the $\widehat{\mathcal{A}}(\widehat{B}_J,\widehat{C}_I )$ denotes any given from of algebraic composition between any two linear maps $\widehat{B}_J$ and $\widehat{C}_I$ in any given Hilbert space. Specifically, this implies that the commutator algebra of the elementary relational operators is exactly inherited from its kinematic counterpart in the form of
\begin{eqnarray}
\label{alg presv}
[\widehat{{P}}_J(t),\widehat{{X}}_I(t)] = [\widehat{{P}}_J,\widehat{{X}}_I](t)\,; \,etc.. 
\end{eqnarray}
Also, the isometry construction implies that each self-adjoint operator in $\mathbb{K}_{t}$ we have 
\begin{eqnarray}
\widehat{\mathcal{O}}=\widehat{\mathcal{O}}^\dagger: \mathbb{S}_{t}\mapsto\mathbb{S}_{t}\Leftrightarrow \widehat{\mathcal{O}}(t)=\widehat{\mathcal{O}}(t)^\dagger: \mathbb{D}_{t}\mapsto\mathbb{D}_{t}.
\end{eqnarray}
Further, the isometry implies that every eigen or coherent  $\widehat{\mathcal{X}}_I$-basis expansion for $\mathbb{K}_t$ induces the associated $\widehat{\mathcal{X}}_I(t)$-basis expansion for $\mathbb{H}_t$ over the same spectrum with the same measure. Explicitly, for any $\psi \in \mathbb{S}_t$ and $|\Psi)\equiv\bra{\psi}\widehat{\mathbb{P}}$ we have
\begin{eqnarray}
\ket{\psi}=\int D\mathcal{X}_I\, \psi(\mathcal{X}_I) \,\ket{T_\mu(t),\mathcal{X}_I}\,\nonumber\\
 \Leftrightarrow \,|\Psi)=\int D\mathcal{X}_I\, \psi(\mathcal{X}_I)\,|\mathcal{X}_I(t))\,;\,\,
|\mathcal{X}_I(t)) \equiv \widehat{I}_t \ket{T_\mu(t),\mathcal{X}_I},
\end{eqnarray}
 and the isometric inner products are encoded in the identified basis dual action of the form
$$\braket{T_\mu(t),\mathcal{X}_I'|T_\mu(t),\mathcal{X}_I}=(\mathcal{X}'_I(t)|\mathcal{X}_I(t)). $$

Lastly, given a complete set of commuting self-adjoint operators $\{\widehat{\mathcal{O}}_I\}$ for $\mathbb{K}_{t_1}$, each state $\Psi\in \mathbb{H}_{\scriptscriptstyle{t_1}}$ can be represented by the normalized ``Dirac wavefunction" over the corresponding complete set of commuting ``elementary relational observables” $\{\widehat{\mathcal{O}}_I(t)\}$ at the moment $t$, given by
\begin{eqnarray}
\label{wave func}
 \Psi({\mathcal{O}}_I(t))\equiv ({\mathcal{O}}_I(t)|\Psi) 
\end{eqnarray}
that evolves unitarily in $t\in  [t_1,t_2] $ simply via the transformations between the $t$-dependent generalized orthognormal bases $\{|{\mathcal{O}}_I(t))\}$.

\subsubsection*{Dynamics in Schr\"odinger and Heisenberg pictures}

We may obtain the evolution Hamiltonian associated to a quantum reference frame in the Schr\"odinger or Heisenberg picture. To do that, we use the natural isometry identification map of
\begin{eqnarray}
\widehat{\iota}_t: \mathbb{K}_{t} \xrightarrow[]{\text{isometry}} K \equiv span\{\ket{{\mathcal{X}}_I}\}\,;\,\,\ket{T_\mu(t),{\mathcal{X}}_I}\mapsto \ket{{\mathcal{X}}_I}
\end{eqnarray}
 to define 
\begin{eqnarray}
\label{red}
(\widehat{P}_{t_2t_1}, \widehat{P}_{t_1t_1}^{-1/2}) \equiv (\widehat{\iota}_{t_2} \widehat{\mathbb{P}}_{t_2t_1}\widehat{\iota}^{\,-1}_{t_1}\,,\,\widehat{\iota}_{t_1} \widehat{\mathbb{P}}_{t_1t_1}^{-1/2}\,\widehat{\iota}^{\,-1}_{t_1})\,: \,S\to S 
\end{eqnarray}
where $S\subset K$ is the dense image of $\mathbb{S}_{t}$ under $\widehat{\iota}_t$. Note that, via the above identification the elementary set for the dynamical sector becomes $\{\widehat{{X}}_I ,\widehat{{P}}_I\}: S\to S$ as a elementary set for $S$. The Dirac wave function in \eqref{wave func} naturally specify a time dependant Schr\"odinger state $\ket{\Psi_{(t)}}\in K$ defined by $$\braket{\mathcal{O}_I|\Psi_{(t_1)}}\equiv \Psi_{(t_1)}(\mathcal{O}_I) \equiv  \Psi({\mathcal{O}}_I(t)).$$
By construction, we have the evolution of the Schr\"odinger state given by
\begin{eqnarray}
\label{eq}
\ket{\Psi_{(t_2)}} =\widehat{U}_{t_2t_1} \ket{\Psi_{(t_1)}}
\end{eqnarray}
where the unitary propagator $\widehat{U}_{t_2t_1}$ is given in the form of
\begin{eqnarray}
\label{prop}
 \widehat{U}_{t_2t_1}\equiv\widehat{P}^{-1/2}_{t_2t_2}\,\widehat{P}_{t_2t_1}\,\,\widehat{P}^{-1/2}_{t_1t_1}\,
: \,S\to S.
\end{eqnarray}
We may now introduced the associated Heisenberg operators via
\begin{eqnarray}
\widehat{\mathcal{X} }^{\scriptscriptstyle{H}}(t)\equiv \widehat{U}^{-1}_{t\, 0}\, \widehat{\mathcal{X} } \,\widehat{U}_{t\, 0}\,: \, S\to S
\end{eqnarray}
It is then straight forward to write down any dynamical quantity for the Dirac theory in the chosen frame, using its Heisenberg-picture counterpart, via the identification $\widehat{\iota}_t$. For an important example, we may express the n-point functions in the Dirac theory by
\begin{eqnarray}
\label{n point1}
 (\Psi| \widehat{\mathcal{O}}_{1}(t) \, \widehat{\mathcal{O}}_{2}(t) \,...\, \widehat{\mathcal{O}}_{n}(t)|\Psi) \nonumber\\
= \bra{\Psi_{(0)}} \widehat{\mathcal{O} }^{\scriptscriptstyle{H}}_{1}(t)\, \widehat{\mathcal{O}}^{\scriptscriptstyle{H}}_{2}(t) \,...\, \widehat{\mathcal{O}}^{\scriptscriptstyle{H}}_{n}(t)\ket{\Psi_{(0)}} 
\end{eqnarray}
Our goal in this work is then to extract the evolution Hamiltonian $\widehat{H}_{t}$ defined by
\begin{eqnarray}
\label{01}
\widehat{H}_{t_1}\equiv-i\hbar\frac{d}{dt}\,\widehat{U}_{t,\, t_1} |_{t=t_1}: K\to K
\end{eqnarray}
expressed in terms of the elementary set, since the obtained explicit form of $\widehat{H}_{t}(\widehat{X}_{I}, \widehat{P}_{I})$ would allow us to generate the first order dynamics by using
\begin{eqnarray}
\label{heis evol}
\frac{d}{dt}\widehat{\mathcal{X} }^{\scriptscriptstyle{H}}(t) = \frac{1}{i\hbar}\big[\widehat{\mathcal{X}}, \widehat{H}_{t}\big]\, \big|_{(\widehat{X}_{I}, \widehat{P}_{I})\to (\widehat{X}^{\scriptscriptstyle{H}}_{I}(t), \widehat{P}^{\scriptscriptstyle{H}}_{I}(t))}
\end{eqnarray}
for any $\mathcal{X}(\widehat{X}_{I}, \widehat{P}_{I})$.

\subsubsection*{Classsical correspondence and tunneling relaxation for reference frame conditions}

Our approach needs further understanding on how and what it means to choose a valid quantum reference frame satisfying the two conditions \eqref{uni cond} and \eqref{uni cond2}. Looking into the constraint surface consisting the ADM constraint solutions in the phase space, one may define a classical reference frame in canonical GR by the specified $t\mapsto X_\mu(t)$ and the restriction $f_{cl}({ X}_\mu, { P}_\mu)>0$ for the region of the constraint surface where the frame is valid (with the $cl$ labeling the ``classical condition"). This is generically equivalent to specifying the frame via the reference fields $T_{\mu}\equiv \theta(f)X_\mu$ with the instantaneous values set by the functions $T_{\mu}(t)=X_\mu(t)$. 

For giving rise to Hamiltonian dynamics, the above choice has to satisfy two phase-space geometric conditions, directly corresponding to our two conditions for the quantum reference frames. First, each point in the constraint surface satisfying $T_\mu=T_\mu(t_1)$ should belong to a unique constraint orbit in the surface, so the set of all such points associated to $t=t_1$ provides a reduced phase-space parametrization of the corresponding set of orbits; as a clear analogy, the condition \eqref{uni cond} demands the ``quantum reduced phase-space" $\mathbb{S}_{t_1}$ to faithfully represent the ``quantum invariant-orbits" $\mathbb{D}_{t_1}$ without degeneracy. Second, the set of the orbits parametrized by the above reduced phase space must remain the same along the different values for $t_1$, so that the Hamiltonian dynamics along $t$ can be obtained describing the fixed set of spacetimes; this condition in the quantum level is clearly given by the \eqref{uni cond2} stating that the completion of  $\mathbb{D}_{t}$ is independent of $t\in [t_1,t_2]$. However, under the same setting in $t\mapsto X_\mu(t)$, there is a crucial difference between the necessary classical restriction $f_{cl}>0$ and the necessary condition $f>0$ to be quantized for a valid quantum reference frame. Based on our previous studies, the quantum tunneling effects tend to make the condition $f>0$ weaker than the $f_{cl}>0$. 

One simple example is given by a model \cite{16 lqc2} of Friedmann-Robertson-Walker loop quantum cosmology (FRW LQC) \cite{16 lqc}, which is a Dirac theory describing a spatial cell of the homogeneous and isotropic universe, governed by a single scalar constraint with a classical form of $${C}\equiv - V^2\frac{\sin^2 ( \sqrt{\Delta}B )}{\Delta} + \beta\, {p_\phi }^2 \equiv \Omega^2+  \beta\, {p_\phi }^2 $$ with the positive $\beta$ and ${\Delta}\sim\hbar^2$ being constants. Leading to the $U(1)$ setting of \eqref{kin2} for the gravitational sector on a single-vertex graph duel to the cell, the phase space coordinates $\{ X_i,P_i\}$ may be given by both $\{ P, e^{\pm i\sqrt{\Delta}B}, \phi, {p_\phi }\}$ and $\{ \Omega, p_\Omega, \phi, {p_\phi }\}$, where the $P$ and $e^{\pm i\sqrt{\Delta}B}$ are respectively the reduced gravitational $U(1)$ flux and holonomy variables detecting the metric volume $V\equiv|P^{\scriptscriptstyle{3/2}}|$ and the extrinsic curvature $B$ of the cell. In the same phase space, the canonical conjugate pairs may be given by $(\phi, p_\phi )$ of the matter scalar field and the gravitational $(\Omega, p_\Omega)$.  Note that the spatial extrinsic curvature $B$ only appears through the curvature holonomy variables in the scalar constraint, and thereby the minimal coupling FRW cosmology is deformed with the quantum geometric corrections mimicking those given by loop quantum gravity.  Using the $\phi$ field as the clock in the earlier WDW treatments, the quantized model with the quantum constraint $\widehat{C}$ has been shown to give the corrected FRW quantum cosmic evolution, with the big bang replaced by a quantum-geometry induced big bounce. In our previous work \cite{12}, we applied our approach to this model with the physical Hilbert space $\mathbb{H}$ defined by $\widehat{\mathbb{P}}\equiv \delta(\widehat{C})$. We identified two valid frames using either the scalar or the gravity sector as the reference sector, defined by the reference field operators $\widehat{T}\equiv \widehat{\phi}_+$ and $\widehat{T}'\equiv \widehat{V}_+$, such that the $\mathbb{S}_t$ and $\mathbb{S}_\tau$ are respectively given by $\ket{T(t)}= \theta(\widehat{p_\phi }) \ket{\phi(t)}$ and $\ket{T'(\tau)}= \theta(\widehat{\Omega}) \theta({\widehat{\cos}\sqrt{\Delta}B}) \ket{V(\tau)}$ with the assigned monotonic $t\mapsto T(t)$ and $\tau \mapsto T(\tau)$. While the frame $t\mapsto \mathbb{K}_t$ reproduced the big bounce evolution from the earlier methods, the second frame $\tau\mapsto \mathbb{K}_\tau$ gave the previously unobtainable Schr\"odinger theory of the scalar field evolving in the quantum spacetime. 

This is a remarkable demonstration of the frame conditions relaxed by the quantum tunneling. Consider the above two classical conditions for the reference frame with the specified monotonic $\tau\mapsto V(\tau)$ over the period of time interval $[\tau_1,\tau_2]$.  It can be checked that first condition would necessarily require $\cos\sqrt{\Delta}B>0$ and $\Omega>0$, and the second condition would demand a further restriction of $|\beta\,p_\phi |<  V(\tau_1)$. That is, the reference frame condition $f_{cl}>0$ in this case must contain all of the three restrictions, and thus the Hamiltonian dynamics would only be defined in the reduced phase space under the truncation in $p_\phi$-- with the truncated space vanishing if $V(\tau_1)$ tends to zero. In the usual methods, this signals a serious obstacle for the corresponding quantum dynamics. Nevertheless, as can be seen from the reference state $\ket{T'(\tau)}$ above, the Schr\"odinger theory is obtained through the quantum reference frame with the $f>0$ consisting only the first two restrictions without any truncation in $p_\phi $, and yet it correctly recovers the desired semi classical ADM evolution of the $\phi$ field as a part of its unitary evolution over  $[\tau_1,\tau_2]\to (-\infty, +\infty)$.  This is due to the fact that a physical state $\Psi \in \mathbb{H}$ has the wave function $\Psi (V,p_\phi)$ with non-zero magnitudes in the ``classically forbidden regions" with $|\beta\,p_\phi |<  V$, and thereby relaxing the condition \eqref{uni cond2}. In this derived Schr\"odinger theory for the scalar field,  a state peaking at $p_\phi(\tau)=\tilde{p}_\phi$ may first evolve semi classically, and then become purely quantum when it enters the period of time with $ |\tilde{p}_\phi| >  V(\tau)$ -- while the quantum unitarity persists. 

In this sense, we see that while the classical conditions for ADM reference frames are useful guides for specifying the associate quantum reference frames, the conditions $f$ for the quantum reference frames are expected to be weaker than the classical counterpart due to the quantum tunnelings.

\subsubsection*{Reference fields as physical background parameters for relational Schr\"odinger dynamics}

Carry on the above into the quantum level, the exact physical meaning of a quantum reference frame involves the reference fields' quantum nature and the quantum-measurement interpretation for the corresponding relational elementary observables. Using the quantum fields internal to the system for the description of time and the relational evolution is a widely explored idea with rich developments \cite{160,161,162}, and let us position our approach in this context. 

First, note that the reference-field operators restricted to their eigenspace $\mathbb{K}_{t}$ are by definition the $c$-number operators of multiplication by ${T}_\mu(t)$, therefore we have $$\widehat{T}_\mu(t)\,:\, \mathbb{H}_{t_1}\to\mathbb{H}_{t_1}\,;\,\, \Psi\mapsto {T}_\mu(t)\cdot\,\Psi$$ as one of the elementary relational observables defined in \eqref{obsv0}, representing the reference fields' values at the moment $t$ as the set of $c$-number operators. This is saying that the reference fields indeed take the values of ${T}_\mu(t)$ at any moment $t$ with zero uncertainties, in the reference frame themselves define. 

However, it is also clear that the $\widehat{T}_\mu(t)$ is the only observable of reference-sector origin, since any reference-sector operator not proportional to $\widehat{T}_\mu$ fails to preserve $\mathbb{K}_{t}$ and can't be pushed forward into a relational observable in $\mathbb{H}_{t_1}$. In this sense, the reference fields serve as a set of (time-dependent) ``external parameters" appearing in the evolution Hamiltonian operator $\widehat{H}_{t}$ specific to the frame-- rather than the quantum dynamical clocks studied in many approaches \cite{160,161,162}. Thus, the existence of a quantum reference frame over an unbounded period \cite{12,13} of time does not require the reference field operators to be canonically conjugate to the ``clock evolution Hamiltonian" with a problematic spectrum unbounded from below. Although all the fields in $\mathbb{K}$ are interacting quantum fields according to the imposed quantum constraints, the quantum reference frame identifies the constrained (physical) quantum degrees of freedom with the space $\mathbb{K}_{t}\subset\mathbb{K}$ in relation to the factorized reference-field eigenstate defining the moment $t$, thereby ``squeezing" all the underlying quantum dynamics into just that for the dynamical sector.

An illustrative example is again given by the FRW LQC model mentioned above. In the quantum reference frame $t\mapsto \mathbb{K}_t$, there is no quantum dynamics for the reference sector of the $\phi$ field, and the reference field $\widehat{\phi}_+(t)= {\phi}_+(t)\cdot$ is effectively a classical background for the Schr\"odinger dynamics of the gravitation sector. Further, as it is shown in \cite{12}, a physical state $\Psi\in \mathbb{D}_{t_1}\cap \mathbb{D}_{\tau_1}$ is also describable in the second frame $\tau\mapsto \mathbb{K}_\tau$ as the Schr\"odinger evolution of the quantum dynamical $\phi$ field governed by the evolution Hamiltonian operator bounded from below. Crucially, via the quantum reference frame transformation, the semi classical quantum gravity state in the background of ${\phi}_+(t)$ is shown to transform into the semi classical state of the quantum scalar field in the background of ${V}_+(\tau)$, with the classically consistent peaked values for the combined dynamical and reference sectors. This points out a possible way of relating the two coexisting roles of the same $\phi$ field-- the one as an``external" reference field abiding to Pauli's view \cite{160,161} or the other as a non-ideal quantum clock field (with dispersion)-- in describing the same physical quantum state $\Psi$.

In a closely parallel consideration, we may also study the physical meaning of the quantum reference frames by searching for a viable quantum measurement interpretation to the elementary relational operators $\{\widehat{X}_I(t), \widehat{P}_I(t))\}$. For instance, under the so-called consistent history interpretations \cite{163,164}, each quantum reference frame for a physical state $\Psi$ may be associated with a complete set of ``histories", each as a coarse-grained quantum trajectory is given by a sequence of projections on to the specified spectral intervals of the Heisenberg-observable bases $\{|{\mathcal O}_n(t_n)\,)\}$ at the moments $t_n$. While the sum of all of the histories is constructed to give a resolution of unity for the Heisenberg state $\Psi$, the interpretation further identifies \cite{163,164} the proper sequences of the projections as the ``consistent histories” exhibiting the desired quantum decoherence we experienced. In the just described application to Dirac quantum gravity, such consistent histories must come with a proper sequence of bases of the elementary relational observables, which would be tied to the proper choices of the quantum reference frame. In this way, we may try to characterize preferred quantum reference frames by their naturalness for the consistent histories to emerge.

We leave those important and interesting questions for our future studies.

\section{Wigner-Weyl representation }

We now move on to compute the form of the desired $\widehat{H}_{t}$ in terms of only the quantum master constraint $\widehat{M}$ and the quantum refrence frame conditions. We will do so by applying the Wigner-Wyel representation \cite{14 wigner} to the construction prescribed in the previous section.

Let us assume a general Hilbert space $\mathbb{K}$ spanned by the eigen or coherent bases of the conjugate pair of complete sets $\{\widehat{x}_i\}$ and $\{\widehat{p}_i\}$, with the resolutions of unity 
\begin{eqnarray}
\label{resol}
\int D{p}_i \ket{{p}_i}\bra{{p}_i}= \int D{x}_i \ket{{x}_i}\bra{{x}_i}= \widehat{I}_{\mathbb{K}}: \mathbb{K}\to \mathbb{K},
\end{eqnarray}
with the integrals taken over the eigen or coherent spectra which are identical to their corresponding ranges in the classical phase space. Using only the data of the spectral matrix elements $\bra{{x}'_i}\widehat{A}\ket{{x}_i}$ and the bases’ dual action $\braket{{p}_i|{x}_i}$, the Wigner-Weyl approach represents any operator $\widehat{A}$ densely defined in $\mathbb{K}$ with a unique phasae space function via the transformation
\begin{eqnarray}
\label{wigtransf}
\widehat{A}\,\xrightarrow[\braket{{p}_i|{x}_i}]{\bra{{x}'_i}\widehat{A}\ket{{x}_i}}\,G_{\scriptscriptstyle{\widehat{A}}}({x}_i,{p}_i).
\end{eqnarray}
The obtained phase space function $G_{\scriptscriptstyle{\widehat{A}}}(\widehat{x}_i,\widehat{p}_i)$ , called the `` symbol" of the $\widehat{A}$, is required to satisfy the following conditions:
\begin{enumerate}
\item The symbols for the elementary operators from $\{\widehat{x}_i,\widehat{p}_i\}$ are simply their corresponding phase space coordinates, so that  
\begin{eqnarray}
\label{prod000}
G_{\scriptscriptstyle{\widehat{p_1}}}({p}_j,{x}_j)={p}_1\,\,; \,\,G_{\scriptscriptstyle{\widehat{x_1}}}({p}_j,{x}_j) ={x}_1.
\end{eqnarray}
\item The adjoint of the operator $\widehat{A}$ is represented by the complex-conjugation, so that
\begin{eqnarray}
\label{prod00}
G_{\widehat{A}^\dagger}=G^*_{\widehat{A}}
\end{eqnarray}
and thus the self-adjoint operators's symbols are always real-valued. 
\item The algebraic product of the operators should be represented by the non-commutative $\star$-product between the corresponding symbols in the form 
\begin{eqnarray}
\label{prod0}
G_{\widehat{A}\widehat{B}}\equiv G_{{\widehat{A}}}\star G_{\widehat{B}}= G_{{\widehat{A}}} \,G_{\widehat{B}}+O(\hbar) \,,
\end{eqnarray}
such that the Poisson brackets of the phase space is reproduced with the $\star$-bracket defined as
\begin{eqnarray}
\label{prod1}
\{G_{{\widehat{A}}}, G_{\widehat{B}} \}_{\star}\nonumber\\
\equiv (G_{{\widehat{A}}}\star G_{\widehat{B}}- G_{\widehat{B}}\star G_{{\widehat{A}}})/i\hbar=G_{[\widehat{A},\,\widehat{B}]/i\hbar}.
\end{eqnarray}
\item For any two vectors $\psi',\psi\in\mathbb{K}$ and their projection $$\ket{\psi'}\bra{\psi}:  \ \mathbb{K}\to \mathbb{K}\,,\,\, \ket{\tilde\psi}\mapsto \braket{\psi|\tilde\psi }\, \ket{\psi'}$$
we have 
\begin{eqnarray}
\label{prod2}
\bra{\psi'}\widehat{A}\ket{\psi}=\label{Tr}
Tr\big[G_{\ket{\psi'}\bra{\psi}} G_{{\widehat{A}}} \big]\nonumber\\
\equiv\int Dp_i\int Dx_i \,G_{\ket{\psi'}\bra{\psi}}({p}_i,\, {x}_i)\,G_{{\widehat{A}}}({p}_i,\, {x}_i).
\end{eqnarray}
\end{enumerate}
In achieving the above, the Wigner-Weyl representation enables us to formulate a quantum theory using purely the phase space functions of the symbols with their non-commutative $\star$-product algebra.

\subsubsection*{Cases of canonical quantization}
For the cases based on the elementary commutators of \eqref{kin1}, here with the elementary set of self-adjoint operators $\{\widehat{x}_i,\widehat{p}_i\}$ having the only nonvanishing commutators of the form $[\widehat{x}_i, \widehat{p}_i]=i\hbar$, the Wigner-Weyl representation is well developed \cite{14 wigner} and understood. Here the eigenspectra for $\widehat{x}_i$ and $\widehat{p}_i$ are identical to the real line $\mathbb{R}$, and we have the transformation \eqref{wigtransf} given by
\begin{eqnarray}
\label{defG0}
G_{\scriptscriptstyle{\widehat{A}}}(\,{p}_i,\, {x}_i\,) \equiv \int D{x}'_{i} \,\,e^{-i\sum_k{p}_k{x}'_k/\hbar} \bra{{x}_i+{x}'_{i}}\widehat{A}\ket{{x}_i-{x}'_{i}}.
\end{eqnarray}
As it is well known, when $\widehat A$ takes a polynomial form of the elementary operators, the function $G_{\scriptscriptstyle{\widehat{A}}}({p}_i,{x}_i)$ is simply given by the form of the operator $\widehat A$ under the totally-symmetric-oredring for the elementary operators. The $\star$-product in the canonical cases is called the Moyal product and can be expressed as 
\begin{eqnarray}
\label{prod}
G_{\widehat{A}\widehat{B}}\equiv G_{{\widehat{A}}}\star G_{\widehat{B}}=\,G_{\scriptscriptstyle{\widehat{A}}}\, e^{{i\hbar}\sum_j( \overleftarrow{\partial_{{x}_j}}\,\,\overrightarrow{\partial_{{p}_j}} - \overleftarrow{\partial_{{p}_j}}\,\,\overrightarrow{\partial_{{x}_j}})/2} \,G_{\widehat{B}}
\end{eqnarray}
where each differentiation acts on either $ G_{\widehat{A}}$ or $ G_{\widehat{B}}$ following the arrows above. Clearly, the $\star$-bracket \eqref{prod1} here indeed gives the Poisson bracket of the symbols in the lowest order of $\hbar$ followed by the quantum corrections.

\subsubsection*{Cases of flux-holonomy quantization}

Currently, the most developed quantum gravity theory based on the elementary commutators of \eqref{kin2} is loop quantum gravity (LQG) \cite{8.2 lqg,15.3} based on a single given graph, with its symmetry reduced cosmological models called the loop quantum cosmology (LQC) \cite{16 lqc,17.1,17.3}. When restricted to a single graph, LQG describes the metric and external curvature of space respectively with the $\mathfrak{su}(2)$-flux and ${SU}(2)$- holonomy variables associated to the graph's edges. Due to the symmetry reductions, the LQC models are usually defined on a single vertex (for the homogeneous models) or a one-dimentional graph (for the spherically symmetric models), while descibed by the reduced $\mathfrak{u}(1)$-flux and ${U}(1)$-holonomy variables. The classical flux-holonomy phase spaces \cite{18} for these theories are well-understood, with their sympletic structures encoded in the flux-holonomy Poisson brackets corresponding to the commutators $\eqref{kin2}$.

Tailored to the flux-holonomy phase space, the Wigner-Weyl representation satisfying \eqref{prod000}-\eqref{prod2} has been constructed \cite{18.1} for the LQC models with the elementary $\mathfrak{u}(1)$- flux operators $\{\widehat{p}_i\}$ and the $U(1)$-holonomy operators $\{\widehat{x}_i\}$, under its own correspondence \eqref{wigtransf} and its own $\star$-product deformed from the Moyal product above. For the $SU(2)$ cases, we note that the Wigner-Weyl representations for the $\mathfrak{su}(2)$ or ${SU}(2)$ systems have been seperately formulated \cite{18.2,18.3}. Thus, it is our expectation that with all the existing foundations and techniques, the Wigner-Weyl representation for the $SU(2)$ flux-holonomy quantization should be within reach for the full LQG theory on a single graph.

\subsubsection*{Some tools for the following}

Our apperoach of quantum reference frames frequently involves a symbol of the form $G_{\widehat{f_1}\widehat{f_2}...\widehat{f_m}}$ where each $\widehat{f_n}$ is of the form
\begin{eqnarray}
\label{0011}
\widehat{f_1}\equiv {f_1}(\widehat{A}_1)\equiv \lim_{k\to\infty}{f^{\scriptscriptstyle{(k)}}_1}(\widehat{A}_1);\nonumber\\
{f^{\scriptscriptstyle{(k)}}_n}(\widehat{A}_n)\equiv \sum^\infty_N \alpha^{(k,N)}_n (\widehat{A}_n)^N.
\end{eqnarray}
The $\widehat{f_1}$ is thus the $k$-sequence limit of the power series ${f^{\scriptscriptstyle{(k)}}_1}(\widehat{A}_1)$ of one single operator $\widehat{A}_1$. We now introduce a useful series expansion for the symbol $G_{\widehat{f_1}\widehat{f_2}...\widehat{f_m}}$ with the leading ``c-number product" term of $f_1(G_{\widehat{A}_1})f_2(G_{\widehat{A}_2})...f_m(G_{\widehat{A}_m})$ followed by the non-commutative corrections of the various degrees of $\star$-brackets amongst the symbols $\{G_{\widehat{A}_n}\}$. To do that, we first decompose the $\star$ -product into the form
\begin{eqnarray}
\label{0022}
\star\equiv\, \cdot\,+\,*
\end{eqnarray}
with the $\cdot$ denoting the usual c-number product, and then we introduce the ``$*$-contraction" of any ordered array ${G_{\widehat{A}_1}}^{\scriptscriptstyle{n_1}}\,{G_{\widehat{A}_2}}^{\scriptscriptstyle{n_2}}...{G_{\widehat{A}_N}}^{\scriptscriptstyle{n_N}}$, denoted by $\{{G_{\widehat{A}_1}}^{\scriptscriptstyle{n_1}}\,{G_{\widehat{A}_2}}^{\scriptscriptstyle{n_2}}...{G_{\widehat{A}_N}}^{\scriptscriptstyle{n_N}}\}_{ ctr}$ and defined as
\begin{eqnarray}
\label{ctr}
&&\{{G_{\widehat{A}_1}}^{\scriptscriptstyle{n_1}}\,{G_{\widehat{A}_2}}^{\scriptscriptstyle{n_2}}...{G_{\widehat{A}_N}}^{\scriptscriptstyle{n_N}}\}_{ ctr}\nonumber\\
\equiv&&\,\big(^{\scriptscriptstyle{n_N}}...\big(^{\scriptscriptstyle{n_2}}\big(^{\scriptscriptstyle{n_1}}\,1[\,(\cdot+\eta*)\,G_{\widehat{A}_1}\,\big)\,]^{\scriptscriptstyle{n_1}}\,[\,(\cdot+\eta*)\,G_{\widehat{A}_2}\,\big)\,]^{\scriptscriptstyle{n_2}}\nonumber\\
&&\,\,\,\,\,\,\,\,\,\,\,\,\,\,\,\,\,\,\,\,\,\,\,\,\,\,\,\,\,\,\,\,...[\,(\cdot+\eta*)\,G_{\widehat{A}_N}\,\big)\,]^{\scriptscriptstyle{n_N}}\,\,\big|^*_{\eta=1}
\end{eqnarray}
where the operations $(\cdot+\eta*) G_{\widehat{A}_n}$ are introduced one-by-one starting from the very left end of the array, followed by the final step $|^*_{\eta=1}$ of setting $\eta=1$ and removing from the resulted expression all the terms with any $G_{\widehat{A}_n}$ from the array acted upon by only the c-number products. Note that by defninition we have $\{{G_{\widehat{A}_1}}^{\scriptscriptstyle{0}}\,{G_{\widehat{A}_2}}^{\scriptscriptstyle{0}}...{G_{\widehat{A}_N}}^{\scriptscriptstyle{0}}\}_{ ctr}=1$. Also, we have introduced the parameter $\eta=1$ just to tract the order of contractions in our calculations, and we will call a term of $O(\eta^m)$ to be of the ``$m$th contraction order" in such a series expansion. Now, for the introduced $ f_n(G_{\widehat{A}_n})$ we further introduce the phase space function $f_1(G_{\widehat{A}_1})\diamond f_2(G_{\widehat{A}_2})...\diamond f_N(G_{\widehat{A}_N})$ defined as
\begin{eqnarray}
\label{diamond}
&&f_1(G_{\widehat{A}_1})\diamond f_2(G_{\widehat{A}_2})...\diamond f_N(G_{\widehat{A}_N})\nonumber\\
\equiv&&\lim_{k\to\infty}\sum^\infty_{{m}_n=0}\, \frac{\{(G_{\scriptscriptstyle{\widehat{A}_1}})^{{m}_1}(G_{\scriptscriptstyle{\widehat{A}_1}})^{{m}_2}...(G_{\scriptscriptstyle{\widehat{A}_N}})^{{m}_N}\}_{ctr}}{{m}_1!\tilde{m}_2!\,...\,{m}_N!} \nonumber\\
&&\,\,\,\,\,\,\times \,\,\partial^{\,\, {m}_1}_{G_{\widehat{A}_1}}f^{\scriptscriptstyle{(k)}}_1(G_{\widehat{A}_1})\cdot\partial^{\,\,{m}_2}_{G_{\widehat{A}_2}}f^{\scriptscriptstyle{(k)}}_2(G_{\widehat{A}_2})...\cdot\partial^{\,\,{m}_N}_{G_{\widehat{A}_N}}f^{\scriptscriptstyle{(k)}}_N(G_{\widehat{A}_N}),\nonumber\\
\end{eqnarray}
As shown in the Appendix, assuming that
\begin{eqnarray}
\label{path}
\lim_{k\to\infty}f^{\scriptscriptstyle{(k)}}_1(G_{\widehat{A}_1})\star f^{\scriptscriptstyle{(k)}}_2(G_{\widehat{A}_2}) ... \star f^{\scriptscriptstyle{(k)}}_N(G_{\widehat{A}_N})\nonumber\\
= f_1(G_{\widehat{A}_1})\star f_2(G_{\widehat{A}_2}) ... \star f_N(G_{\widehat{A}_N})
\end{eqnarray}
we then have
\begin{eqnarray} 
\label{000}
G_{\widehat{f_1}\widehat{f_2}...\widehat{f_N}}
&&=G_{\widehat{f_1}}\star G_{\widehat{f_2}} ...\star G_{\widehat{f_N}} \nonumber\\
&&=f_1(G_{\widehat{A}_1})\diamond f_2(G_{\widehat{A}_2})...\diamond f_N(G_{\widehat{A}_N}),
\end{eqnarray}
hence the expression \eqref{diamond}, which we will call as the ``diamond expansion" for $G_{\widehat{f_1}\widehat{f_2}...\widehat{f_N}}$, provides the desired series expansion with the non-commutative corrections given in the ``Taylor series form" over all orders of the $*$-contractions. 

Lastly, let us give some important examples for the functions $f_n$ in the above. First, the Dirac-delta distribution and the associated unit-step function can be expressed as
\begin{eqnarray}
\label{delta}
\delta(x)\equiv \lim_{k\to \infty}\delta^{\scriptscriptstyle{(k)}}(x)\equiv \lim_{k\to \infty} \int_{-k}^{k} e^{ik'x}dk' \,;\nonumber\\
\theta(x)\equiv \lim_{k\to \infty}\theta^{\scriptscriptstyle{(k)}}(x)\equiv \lim_{k\to \infty} \int_{-k}^{k}(\pi\delta(k')+\frac{1}{ik'} )e^{ik'x}\,dk'\,.
\end{eqnarray}
Second, the inverse square-root function $x^{-1/2}$ for $x\in(0,\infty)$ can be expressed as 
\begin{eqnarray}
\label{invsqrt}
\frac{1}{\sqrt{x}} \big|_{x\in(0,\infty)}\equiv \lim_{k\to \infty}\frac{1}{\sqrt[(k)]{x}}\equiv\lim_{k\to \infty} \int_{-k}^{k} \sqrt{\frac{\pi}{|k'|}}e^{ik'x}\,dk'.\nonumber\\
\end{eqnarray}

\section{First-integral quantum dynamic in the Weyl-symbol phase space}

Under a valid quantum reference frame satisfying the two conditions \eqref{uni cond} and \eqref{uni cond2}, the unitary Heisenberg evolution \eqref{heis evol} of the elementary relational observables can now be expressed in the $\star$-bracket representation via \eqref{prod1}. This gives the first order flow of all the dynamical quantities, such as the evolution of the n-point function in the form of 
\begin{eqnarray}
\label{gamma2}
(\Psi| \widehat{\mathcal{O}}_{1}(t) \, \widehat{\mathcal{O} }_{2}(t) \,...\, \widehat{\mathcal{O}}_{n}(t)|\Psi) \nonumber\\
=Tr \left[ {G}_{\ket{\Psi_{(0)}}\bra{ \Psi_{(0)}}}\,\,{G}_{\scriptscriptstyle{\widehat{\mathcal{O}}^{\scriptscriptstyle{H}}_{1}(t)}}\star {G}_{\scriptscriptstyle{\widehat{\mathcal{O}}^{\scriptscriptstyle{H}}_{2}(t)}}\star\, ... \,\star {G}_{\scriptscriptstyle{\widehat{\mathcal{O}}^{\scriptscriptstyle{H}}_{n}(t)}} \right]\nonumber\\
\end{eqnarray}
Here we use our established Heisenberg picture on the right for the Dirac theory on the left, and all the observables' symbols ${G}_{\scriptscriptstyle{\widehat{\mathcal{O}}^{\scriptscriptstyle{H}}_{n}(t)}}$ are constructed via \eqref{comp} and \eqref{prod0} from the symbols of the elementary relational operators $\{{G}_{\scriptscriptstyle{\widehat{X}^{\scriptscriptstyle{H}}_{I}(t)}},{G}_{\scriptscriptstyle{\widehat{P}^{\scriptscriptstyle{H}}_{I}(t)}}\} $, which evolves according to
\begin{eqnarray}
\label{evolu1}
\big({G}_{\scriptscriptstyle{\widehat{X}^{\scriptscriptstyle{H}}_{I}(0)}},{G}_{\scriptscriptstyle{\widehat{P}^{\scriptscriptstyle{H}}_{J}(0)}}\big)= (X_I, P_J) \,\,;\nonumber\\
\frac{\partial}{\partial t}\big({G}_{\scriptscriptstyle{\widehat{X}^{\scriptscriptstyle{H}}_{I}(t)}},{G}_{\scriptscriptstyle{\widehat{P}^{\scriptscriptstyle{H}}_{J}(t)}}\big)\nonumber\\
= \left\{\,\big(G_{\scriptscriptstyle{\widehat{X }_{I}}},G_{\scriptscriptstyle{\widehat{P }_{J}}}\big)\,, \,G_{\scriptscriptstyle{\!\widehat{H}_t}} \,\right\}_\star\,|_{(G_{\widehat{X }_{I}},G_{\widehat{P }_{I})}\to(G_{\widehat{X }^{\scriptscriptstyle{H}}_{I}(t)},G_{\widehat{P }^{\scriptscriptstyle{H}}_{I}(t)}) }.\nonumber\\
\end{eqnarray}
In this formulation, the evolution is then determined by the initial quantum state given by the symbol $G_{\ket{\Psi_{(0)}}\bra{ \Psi_{(0)}}}(P_I, X_I)$ and the flow generated by \eqref{evolu1}. The remaining crucial task is now obtaining the quantum physical Hamiltonian $G_{\scriptscriptstyle{\!\widehat{H}_t}}$-- in its Weyl symbol form.

\section{Weyl symbol of the evolution Hamiltonian under a quantum reference frame}

Under a choice of the reference sector, we set $\{\widehat{x}_i\}\equiv \{\widehat{{X}}_\mu,\widehat{{X}}_I\}$ and $\{\widehat{p}_i\}\equiv \{\widehat{P}_\mu, \widehat{P}_I\}$ for the resolution of unity \eqref{resol} used in the Weyl transformation. In this work, we assume the choice is such that the basis $\{\ket{{x}_i}\equiv\ket{{X}_\mu, {X}_I} \equiv \ket{{X}_\mu}\otimes\ket{{X}_I}\}$ is a common eigenbasis of $\{\widehat{{X}}_\mu\}$, with the continuous eigenspectra and the basis dual action given by
\begin{eqnarray}
\label{reff0}
\braket{{X}'_\mu|{X}_\nu}= \delta({X}'_\mu-{X}_\mu)
\end{eqnarray}
where the dirac-delta distribution is with respect to the reference-sector's measure $D{X}_\mu$ in the resolution of unity. Further, with a properly assigned $t\to{X}_\mu(t)$ and $t\to\mathcal{N}_\nu(t)$ we assume the natural correspondence from the classical phase space transformation
\begin{eqnarray}
\partial_t {X}_\mu(t) =\{{X}_\mu\,, \,\mathcal{N}_\nu(t)\,{P}_\nu\}|_{{X}_\mu={X}_\mu(t)} \nonumber\\
\Rightarrow i\hbar\partial_t\ket{{X}_\mu(t),X_I}= \mathcal{N}_\nu(t)\,\widehat{P}_\nu\,\ket{{X}_\mu(t),X_I}
\end{eqnarray}
to the quantum counterpart. Note that the above assumptions hold not only for the obvious canonical cases of \eqref{kin1} when the $\widehat{{X}}_\mu$ and $\widehat{P}_\mu$ are canonically conjugate pairs, but also for the cases of \eqref{kin2} when the $\widehat{{X}}_\mu$ and $\widehat{P}_\mu$ are respectively the holonomy operators and their conjugate flux operators. Restricting to such type of reference sectors, we will focus on a large class of quantum reference frames $t\mapsto\mathbb{K}_t$ characterized by \begin{eqnarray}
\label{reff}
\mathbb{K}_t= span\{\ket{T_\mu(t),X_I} \}=span\{\theta(\widehat{f})\,\ket{{X}_\mu(t),X_I}\}\,\nonumber\\
\end{eqnarray}
with the specified functions $t\mapsto{X}_\mu(t)$ and the given self-adjoint operator $$\widehat{f}=\widehat{f}(\widehat{X}_\nu, \widehat{P}_\nu)=\widehat{f}^\dagger\,:\, \mathbb{S}\to\mathbb{S}$$ quantizing the restriction condition $f(X_{\nu},P_{\nu})>0$ discussed earlier; as mentioned, this class of quantum reference frames have been successfully applied in obtaining the Schr\"odinger dynamics for the loop quantum cosmology models \cite{12,13}. Using \eqref{prop} and \eqref{01} with the frame specified here, we now compute the symbol $G_{\scriptscriptstyle{\!\widehat{H}_t}}$ from the basic building blocks of the associated $G_{\scriptscriptstyle{\widehat{ P}_{t_2,t_1}}}$ which by definition satisfy ($r=0,1$)
\begin{eqnarray}
\label{physH34}
&&(-i\hbar\partial_{t_1})^rTr\!\left[G_{\scriptscriptstyle{\ket{\phi}\bra{\phi'} }}G_{\scriptscriptstyle{\widehat{ P}_{t_2,t_1}}}\right] \,|_{t_1=t_2}\nonumber\\
&=&(-i\hbar\partial_{t_1})^r \bra{\phi'} \widehat{ P}_{t_2,t_1}\ket{\phi}\,|_{t_1=t_2}\nonumber\\
&=&(-i\hbar\partial_{t_1})^r\bra{T_\mu(t_2),\phi'} \widehat{\mathbb{P}}\ket{T_\mu(t_1),\phi }|_{t_1=t_2}\nonumber\\
&=&\mathcal{N}_\nu(t_1)\,\bra{X_\mu(t_1),\phi'}\theta(\widehat{f})\,\delta(\widehat{M})\, \theta(\widehat{f})\widehat{P}^r_{\nu}\ket{ X_\mu(t_1),\phi}\nonumber\\
\end{eqnarray}
where we have introduced
\begin{eqnarray}
\widehat{\iota}_{t_1} \ket{\phi, T_\mu(t_1)} = \ket{\phi} \,\,; \,\,\widehat{\iota}_{t_2} \ket{\phi', T_\mu(t_2)} = \ket{\phi'}
\end{eqnarray}
with any two dynamical-sector vectors $\ket{\phi}, \ket{\phi'}\in S$. Further, since by construction the the reference and dynamical sectors are disentangled in $\mathbb{K}_{t_1}$, we have 
\begin{eqnarray}
\label{dir}
&&G_{\scriptscriptstyle{\ket{\phi, X_\mu(t_1)}\bra{\phi', X_\mu(t_1)} }}(X_i,P_i)\nonumber\\
=&& G_{\scriptscriptstyle{\ket{ X_\mu(t_1)}\bra{ X_\mu(t_1)} }}(X_\nu,P_\nu)\cdot G_{\scriptscriptstyle{\ket{\phi}\bra{\phi'} }}(X_I,P_I)\nonumber\\
\end{eqnarray}
Using this direct product form to express the reult of \eqref{physH34} above in Weyl symbols, we then have
\begin{eqnarray}
\label{physH345}
&&(-i\hbar\partial_{t_1})^rTr\!\left[G_{\scriptscriptstyle{\ket{\phi}\bra{\phi'} }}G_{\scriptscriptstyle{\widehat{ P}_{t_2,t_1}}}\right] \,|_{t_1=t_2}\nonumber\\
&=&\dot{T}_{\nu(t_1)} tr_{\scriptscriptstyle{D}}\!\left[G_{\scriptscriptstyle{\ket{\phi}\bra{\phi'} }}\,tr_{\scriptscriptstyle{R}}\!\left[ \,G_{\scriptscriptstyle{ \ket{{X}_\mu(t_1)}\bra{{X}_\mu(t_1)}} }\,\,G_{\theta(\widehat{f})\,\delta(\widehat{M})\, \theta(\widehat{f}){\widehat{P}^r_{\nu}}}\right]\right]\,,\nonumber\\
\end{eqnarray}
where the total trace $Tr=tr_{\scriptscriptstyle{D}} \,tr_{\scriptscriptstyle{R}}$ is taken by the partial tracing $tr_{\scriptscriptstyle{D}}$ over the dynamical sector and then that $tr_{\scriptscriptstyle{R}}$ over the reference sector. Since due to \eqref{reff0} we have $$ G_{\scriptscriptstyle{\ket{ X_\mu(t_1)}\bra{ X_\mu(t_1)} }}(X_\nu,P_\nu)= \delta(X_\nu-X_\nu(t_1))$$
and \eqref{physH345} holds for all $\ket{\phi}, \ket{\phi'}\in S$, we have now obtained the results of
\begin{eqnarray}
\label{32}
&&(-i\hbar\partial_{t_2})^rG_{\scriptscriptstyle{\widehat{ P}_{t_2,t_1}}}|_{t_1=t_2}\nonumber\\
=&&\mathcal{N}_\nu(t_1)\, tr_{\scriptscriptstyle{R}}\!\left[ \,G_{\scriptscriptstyle{ \ket{{X}_\mu(t_1)}\bra{{X}_\mu(t_1)}} }\,P^r_{\nu}\diamond \theta(G_{\scriptscriptstyle{\widehat{f}}})\diamond \delta(G _{\scriptscriptstyle{\widehat{M}}}) \diamond\theta(G_{\scriptscriptstyle{\widehat{f}}})\,\right]\,\nonumber\\
=&&\mathcal{N}_\nu(t_1)\int \,D{P}_\mu \,\,\, {P}^r_{\nu}\diamond \theta(G_{\scriptscriptstyle{\widehat{f}}})\diamond \delta(G _{\scriptscriptstyle{\widehat{M}}}) \diamond\theta(G_{\scriptscriptstyle{\widehat{f}}})\big|_{X_\mu={X}_\mu(t_1)}
;\nonumber\\
\nonumber\\
&&(i\hbar\partial_{t_1})^rG_{\scriptscriptstyle{\widehat{ P}_{t_2,t_1}}}|_{t_1=t_2}\nonumber\\
=&&\mathcal{N}_\nu(t_1)\, tr_{\scriptscriptstyle{R}}\!\left[ \,G_{\scriptscriptstyle{ \ket{{X}_\mu(t_1)}\bra{{X}_\mu(t_1)}} }\,\theta(G_{\scriptscriptstyle{\widehat{f}}})\diamond \delta(G _{\scriptscriptstyle{\widehat{M}}}) \diamond\theta(G_{\scriptscriptstyle{\widehat{f}}})\diamond P^r_{\nu} \,\right]\,\nonumber\\
=&&\mathcal{N}_\nu(t_1)\int \,D{P}_\mu \,\,\, \theta(G_{\scriptscriptstyle{\widehat{f}}})\diamond \delta(G _{\scriptscriptstyle{\widehat{M}}}) \diamond\theta(G_{\scriptscriptstyle{\widehat{f}}})\diamond{P}^r_{\nu}\big|_{X_\mu={X}_\mu(t_1)}.\nonumber\\
\end{eqnarray}
Here we have replace the $\star$-products with the diamond expansion via \eqref{000} along with \eqref{delta}, and the integral over the $X_\mu$ for the partial trace has been carried out with the $ G_{\scriptscriptstyle{\ket{ X_\mu(t_1)}\bra{ X_\mu(t_1)} }}$ taking the above Dirac distribution form, leaving only the integral over the $P_\mu$ in the final expressions.

Further, the propagator given in \eqref{prop} can now be expressed under the diamond expansion, using $(f_1(x), f_2(x), f_3(x))=(1/\sqrt{x}\,,\, x\,,\, 1/\sqrt{x})$ and \eqref{invsqrt}, in the form of 
\begin{eqnarray}
\label{1}
G_{\scriptscriptstyle{ \widehat{U}_{t_2,\,t_1}}}= \frac{1}{\sqrt{G_{\scriptscriptstyle{\widehat{ P}_{t_2,t_2}}}}}\diamond G_{\scriptscriptstyle{\widehat{ P}_{t_2,\,t_1}}} \diamond\frac{1}{\sqrt{G_{\scriptscriptstyle{\widehat{ P}_{t_1,t_1}}}}} \,,
\end{eqnarray}
from which we obtain
\begin{eqnarray}
\label{2}
&&G_{\scriptscriptstyle{\!\widehat{H}_t}}\nonumber\\
&=&-i\hbar(\partial_{t_2} -\partial_{t_1})\frac{1}{\sqrt{G_{\scriptscriptstyle{\widehat{ P}_{t_2,t_2}}}}}\diamond G_{\scriptscriptstyle{\widehat{ P}_{t_2,\,t_1}}} \diamond\frac{1}{\sqrt{G_{\scriptscriptstyle{\widehat{ P}_{t_1,t_1}}}}} \bigg|_{t_1=t_2=t}
\end{eqnarray}

Finally, we have derived the the desired set of closed formulae \eqref{2} and \eqref{32} which gives the evolution Hamiltonian $G_{\scriptscriptstyle{\!\widehat{H}_t}}$ in terms of only the governing quantum master constraint $G_{\widehat{M}}$ and the quantum-reference-frame conditions $\{G_{\scriptscriptstyle{\widehat{f}}}\,, \,G_{ \scriptscriptstyle{\ket{{X}_\mu(t)}\bra{{X}_\mu(t)}} }\}$.

\section{Preliminary series-expansion scheme for $G_{\scriptscriptstyle{\!\widehat{H}_t}}$ }

\subsection{Series expansion over contractions $\eta^N$ and Fourier modes $k$} 

Our construction gives the result \eqref{2} in a natural series expansion expression, over the Fourier modes $k$ introduced in \eqref{delta} and \eqref{invsqrt} and the contractions of degree $\eta^N$ introduced in \eqref{2} and \eqref{32} acting amongst the power series of $G_{\widehat{M}}$ and $G_{\scriptscriptstyle{\widehat{f}}}$. To see this, we use \eqref{delta} and express $G_{\scriptscriptstyle{\widehat{ P}_{t_2,t_1}}}$ as
$$G_{\scriptscriptstyle{\widehat{ P}_{t_2,t_1}}}=\lim_{k\to \infty}{}^{\scriptscriptstyle{k}}G_{\scriptscriptstyle{\widehat{ P}_{t_2,t_1}}}$$ where following the definition \eqref{diamond} we introduce ($r=0,1$)
\begin{eqnarray}
\label{33}
&&(\frac{i\hbar}{\mathcal{N}_\nu(t_1)\,}\,\partial_{t_2})^r\,\,\,{}^{\scriptscriptstyle{k}}G_{\scriptscriptstyle{\widehat{ P}_{t_2,t_1}}}|_{t_1=t_2}\nonumber\\
\equiv&&\int \,D{P}_\mu \,\,\, \widehat{P}^r_{\nu}\diamond \theta^{\scriptscriptstyle{(k)}}(G_{\scriptscriptstyle{\widehat{f}}})\diamond \delta^{\scriptscriptstyle{(k)}}(G _{\scriptscriptstyle{\widehat{M}}}) \diamond\theta^{\scriptscriptstyle{(k)}}(G_{\scriptscriptstyle{\widehat{f}}})\big|_{X_\mu={X}_\mu(t_1)}
;\nonumber\\
\nonumber\\
&&(\frac{-i\hbar}{\mathcal{N}_\nu(t_2)\,}\partial_{t_1})^r\,\,\,{}^{\scriptscriptstyle{k}}G_{\scriptscriptstyle{\widehat{ P}_{t_2,t_1}}}|_{t_1=t_2}\nonumber\\
\equiv&&\int \,D{P}_\mu \,\,\, \theta^{\scriptscriptstyle{(k)}}(G_{\scriptscriptstyle{\widehat{f}}})\diamond \delta^{\scriptscriptstyle{(k)}}(G _{\scriptscriptstyle{\widehat{M}}}) \diamond\theta^{\scriptscriptstyle{(k)}}(G_{\scriptscriptstyle{\widehat{f}}})\diamond\widehat{P}^r_{\nu}\big|_{X_\mu={X}_\mu(t_1)}\,.\nonumber\\
\end{eqnarray}
Further, since ${}^{\scriptscriptstyle{k}}G_{\scriptscriptstyle{\widehat{ P}_{t_2,t_1}}}$ is a power series over $\eta^N$ by the diamond expansion, we have the series expansion in the form of 
\begin{eqnarray}
\label{5}
G_{\scriptscriptstyle{\widehat{ P}_{t_1,t_2}}}\equiv \lim_{k\to\infty} \sum_{N=1}^\infty \eta^{N}\, {}^{\scriptscriptstyle{k}}G^{\scriptscriptstyle{(N)}}_{\scriptscriptstyle{{\widehat{ P}_{t_2,t_1}}}}\equiv \sum_{k=1}^\infty \sum_{N=1}^\infty \eta^{N}\, {\Delta}^{k}G^{\scriptscriptstyle{(N)}}_{\scriptscriptstyle{{\widehat{ P}_{t_2,t_1}}}} |_{\eta=1}. \nonumber\\
\end{eqnarray}
where we express the contributions from the unit-interval Fourier modes with the notation $$ {\Delta}^{k}G_{\scriptscriptstyle{{\widehat{A}}}}\equiv {}^{\scriptscriptstyle{k+1}}G_{\scriptscriptstyle{{\widehat{A}}}}- {}^{\scriptscriptstyle{k}}G_{\scriptscriptstyle{{\widehat{A}}}}.$$
Next, in the same manner and using \eqref{invsqrt} we have
\begin{eqnarray}
G_{\scriptscriptstyle{\!\widehat{H}_t}}=\lim_{k\to \infty} {}^{\scriptscriptstyle{k}}G_{\scriptscriptstyle{\!\widehat{H}_t}}
\end{eqnarray}
which for any given $k$ is defined by the diamond expansion with $(f_1(x), f_2(x), f_3(x))=(1/\!\sqrt[(k)]{x}\,,\, x\,,\,1/\!\sqrt[(k)]{x})$ as
\begin{eqnarray}
\label{7}
&&{}^{\scriptscriptstyle{k}}G_{\scriptscriptstyle{\!\widehat{H}_t}}\nonumber\\
&&\equiv i\hbar(\partial_{t_1} -\partial_{t_2})\frac{1}{\sqrt[(k)]{{}^{\scriptscriptstyle{k}}G_{\scriptscriptstyle{\widehat{ P}_{t_2,t_2}}}}}\diamond {}^{\scriptscriptstyle{k}}G_{\scriptscriptstyle{\widehat{ P}_{t_2,\,t_1}}} \diamond\frac{1}{\sqrt[(k)]{{}^{\scriptscriptstyle{k}}G_{\scriptscriptstyle{\widehat{ P}_{t_1,t_1}}}}} \bigg|_{t_1=t_2=t}.\nonumber\\
\end{eqnarray}
The above, involving another round of contractions, is again a power series in $\eta^N$ leading to the series expansion of 
\begin{eqnarray}
\label{8}
G_{\scriptscriptstyle{\!\widehat{H}_t}}\equiv \lim_{k\to\infty} \sum_{N=1}^\infty \eta^{N}\, {}^{\scriptscriptstyle{k}}G^{\scriptscriptstyle{(N)}}_{\scriptscriptstyle{\!\widehat{H}_t}}\equiv \sum_{k=1}^\infty \sum_{N=1}^\infty \eta^{N}\, {\Delta}^{k}G^{\scriptscriptstyle{(N)}}_{\scriptscriptstyle{\!\widehat{H}_t}} |_{\eta=1}. \nonumber\\
\end{eqnarray}
Here the $N$ is the total degree of contractions from both of the diamond expansions involved in \eqref{7} and \eqref{33}. The above thus provides a tentative version of the desired series expansion for $G_{\scriptscriptstyle{\!\widehat{H}_t}}$ over the $\eta^N$ and $k$. 

On the convergence of the expansions \eqref{8} and \eqref{5}, we observe the following. The convergence in $k\to \infty$ is expected from \eqref{33} and \eqref{7}, since the self-adjoint operators $\widehat{M}$, $\widehat{f}$ and $\widehat{ P}_{t,\,t}$ may be treated as the multiplicative real numbers of their spectra, with the values lying in the domains of the functions \eqref{delta} and \eqref{invsqrt} converging in the $k\to \infty$ limits. The convergence in $N\to \infty$ further rests on the validity condition \eqref{path} for the diamond expansion, and when given, it always happens with a supressing power-series expansion factor of $\hbar^N$.

\subsection{ Further expansion of $G_{\scriptscriptstyle{\!\widehat{H}_t}}$ for semi classical limits} 

When both the quantum master constraint and the quantum frame conditions have classical counterparts, there must be associated power series expansions for their symbols over $\hbar$ in the form
\begin{eqnarray}
\label{semi1}
(G_{\scriptscriptstyle{\widehat{M}}}, G_{\scriptscriptstyle{\widehat{f}}})\equiv \sum_{r=0}^{\infty} \lambda^r(G^{\scriptscriptstyle{(r)}}_{\scriptscriptstyle{\widehat{M}}}, G^{\scriptscriptstyle{(r)}}_{\scriptscriptstyle{\widehat{f}}})\big|_{\lambda=1}\sim \hbar^r\,;\nonumber\\
\, (G^{\scriptscriptstyle{(0)}}_{\scriptscriptstyle{\widehat{M}}}, G^{\scriptscriptstyle{(0)}}_{\scriptscriptstyle{\widehat{f}}})\equiv (\,\sqrt{\sum_\mu {\bar C}^2_{\mu}}\,,\,\bar{f}\,).
\end{eqnarray}
where the $G^{\scriptscriptstyle{(0)}}_{\scriptscriptstyle{\widehat{M}}}$ is given by the set of real phase space functions $\{\bar{C}_{\mu}\}$ which may be deformed from the ADM constraints $\{ {C}_{\mu}\}$. Here we again insert the $\lambda=1$ for the power tracking. Then, by inserting the above into the previous expansion \eqref{8}, we obtain the further expansion of $G_{\scriptscriptstyle{\!\widehat{H}_t}}$ in the form of 
\begin{eqnarray}
\label{87}
G_{\scriptscriptstyle{\!\widehat{H}_t}}\equiv \lim_{k\to \infty} \sum_{N=1}^\infty \eta^{N}\, {}^{\scriptscriptstyle{k}}G^{\scriptscriptstyle{(N)}}_{\scriptscriptstyle{\!\widehat{H}_t}} |_{\eta=1}\nonumber\\
\equiv\lim_{k\to \infty} \sum_{N=1}^\infty \sum_{r=0}^\infty\eta^{N}\lambda^r\, {}^{\scriptscriptstyle{k}}G^{\scriptscriptstyle{(N,r)}}_{\scriptscriptstyle{\!\widehat{H}_t}} |_{\eta,\lambda=1}\nonumber\\
\equiv \sum_{k=1}^\infty \sum_{N=1}^\infty\sum_{r=0}^\infty \eta^{N}\lambda^r\, \,{\Delta}^{k}G^{\scriptscriptstyle{(N,r)}}_{\scriptscriptstyle{\!\widehat{H}_t}} |_{\eta,\lambda=1}\sim \hbar^{N+r} . 
\end{eqnarray}
which suggests a semi-classical limit of classical Hamiltonian dynamics ``governed" by the constraints $\{\bar{C}^2_{\mu}\}$ under the reference frame conditions of $t\mapsto {X}_\mu(t)$ and $\bar{f}>0$. Clearly, this limit can exist only in the region of the space of $(X_I,P_I,t)$ where the limit $\lim_{k\to\infty}{}^{\scriptscriptstyle{k}}G^{\scriptscriptstyle{(0,0)}}_{\scriptscriptstyle{\!\widehat{H}_t}}(X_I,P_I)$ converges to a non-vanishing leading term. 

Indeed, as we now demonstrate, for the given decomposition \eqref{semi1} one could identify a region in the space of $(X_I,P_I,t)$ in which the limit ${}^{\scriptscriptstyle\infty}G^{\scriptscriptstyle{(0,0)}}_{\scriptscriptstyle{\!\widehat{H}_t}}(X_I,P_I)$ exists and reproduces the familiar evolution Hamiltonian under the specified frame.

The $(N=0, r=0)$-order contribution to \eqref{33} can be evaluated with all the diamond products replaced with just the c-number product. This leads to the expressions of ($s=0,1$)
\begin{eqnarray}
\label{3}
&&(-i\hbar\partial_{t_2})^s\,\,{}^{\scriptscriptstyle\infty}G^{\scriptscriptstyle{(0,0)}}_{\scriptscriptstyle{{\widehat{ P}_{t_2,t_1}}}}|_{t_1=t_2} = (i\hbar\partial_{t_1})^s\,\,{}^{\scriptscriptstyle\infty}G^{\scriptscriptstyle{(0,0)}}_{\scriptscriptstyle{{\widehat{ P}_{t_2,t_1}}}}|_{t_1=t_2} \nonumber\\
=&&\mathcal{N}^s_\nu(t_1)\int \,D{P}_\mu \,\,\, {P}^s_{\nu}\cdot \theta(\bar{f})^2\cdot \delta\big(\,\sqrt{\sum_\mu {\bar C}^2_{\mu}}\,\big) \,\big|_{X_\mu={X}_\mu(t_1)}
\nonumber\\
\end{eqnarray}
Importantly, the functions ${\bar C}_{\mu}$ and the integration variables ${P}_\mu$ are related in their measures by the Jaccobian 
\begin{eqnarray}
\label{det}
&&J (X_I, P_I, {P}_\mu, t)
\equiv \text{det}\left[ \frac{\partial {\bar C}_{\mu'}}{\partial {P}_{\nu'}}\right](X_I, P_I, {X}_\nu, {P}_\mu )\big|_{ {X}_\nu=T_\mu(t)} \,.\nonumber\\
\end{eqnarray}
In the special case when the set $\{{\bar C}_{\mu}\}$ heppens to form a first class constraint system, one may recognized this quantity as the Fadeev-Poppov determinant, if the specification of the frame is treated as a gauge-fixing for the symmetry generated by the system. 

For each decomposition \eqref{semi1} and quantum reference frame, let us define the associated ``semi classical region" $\Omega_{s.cl.}$ in the space of $({ X}_I, { P}_I, t)$ to be such that: (1) for each $(X_I,P_I,t)\in\Omega_{s.cl.}$ there is a non-empty and countable set of solutions $\{\,{P}_{\mu,i}({ X}_I, { P}_I, t)\,\}$ labeled by the index $i$, consisting of the maximal set of $P_\mu$ values satisfying
\begin{eqnarray}
\bar{C}_\mu (X_I, P_I, {P}_{\mu,i}, X_\mu(t))\, \restriction_{\Omega_{s.cl.}}=0\,\,;\,\,\bar{f}({P}_{\mu,i}, X_\mu(t))\, \restriction_{\Omega_{s.cl.}} >0;\nonumber
\end{eqnarray}
(2) the values of the Jaccobian $\,J_i(X_I,P_I,t)\equiv J (X_I, P_I, {P}_{\mu,i}, t)$ evaluated at the solutions satisfy $$0<\sum_i \big|J_i^{-1}(X_I,P_I,t) \big| \,\restriction_{\Omega_{s.cl.}}<\infty,$$ which in the case of the first-class system is just the familiar regularity condition for the gauge-fixing. Indeed, the integral \eqref{3} is finite and non-vanishing precisely in the region $\Omega_{s.cl.}$ with the result easily obtained as
\begin{eqnarray}
\label{31}
(-i\hbar\partial_{t_2})^s\,{}^{\scriptscriptstyle\infty}G^{\scriptscriptstyle{(0,0)}}_{\scriptscriptstyle{{\widehat{ P}_{t_2,t_1}}}}\,\restriction_{t_1=t_2\,,\, \Omega_{s.cl.}}= \,\sum_i \,\,\big|J^{-1}_i\big| \,\,({{P}_{\mu,i}}\,\mathcal{N}_\mu(t))^s\,;\nonumber\\
\end{eqnarray}
Finally, since the above with $s=0$ is non-vanishing in the region $\Omega_{s.cl.}$, the $k\to\infty$ limit for just the $(0,0)$-order contribution in \eqref{7} is itself well-defined and gives
\begin{eqnarray}
\label{intform}
&&{}^{\scriptscriptstyle\infty}G^{\scriptscriptstyle{(0,0)}}_{\scriptscriptstyle{\!\widehat{H}_t}}\,\restriction_{\Omega_{scl}}\nonumber\\
&&=i\hbar(\partial_{t_1} -\partial_{t_2})\frac{1}{\sqrt[\scriptscriptstyle{(\infty)}]{{}^{\scriptscriptstyle{\infty}}G^{\scriptscriptstyle{(0,0)}}_{\scriptscriptstyle{\widehat{ P}_{t_2,t_2}}}}} \frac{1}{\sqrt[\scriptscriptstyle{(\infty)}]{{}^{\scriptscriptstyle{\infty}}G^{\scriptscriptstyle{(0,0)}}_{\scriptscriptstyle{\widehat{ P}_{t_1,t_1}}}}}{}^{\scriptscriptstyle{\infty}}G^{\scriptscriptstyle{(0,0)}}_{\scriptscriptstyle{\widehat{ P}_{t_2,\,t_1}}}\, \restriction_{t_1=t_2=t}\nonumber\\
&&=\big(\,\sum_j \,\big|J^{-1}_j\big|\,\,\big)^{-1} \big(\,\sum_i \,\,\big|J^{-1}_i \big|\,\,{P}_{\mu,i}\,\mathcal{N}_\mu(t)\,\big)\,. \nonumber\\
\end{eqnarray}
Moreover, when there is a "classical region" $\Omega_{cl}\subset\Omega_{scl}$ where the set $\{{P}_{\mu,i}\}$ contains a unique element ${P}_{\mu,1}$ we then have 
\begin{eqnarray}
\label{intform2}
{}^{\scriptscriptstyle\infty}G^{\scriptscriptstyle{(0,0)}}_{\scriptscriptstyle{\!\widehat{H}_t}} \restriction_{\Omega_{cl}}
={P}_{\mu,1}\,\mathcal{N}_\mu(t)\,. \nonumber\\
\end{eqnarray}
Again, when the set $\{{\bar C}_{\mu}\}$ happens to form a first class constraint system, the above indeed is the familiar evolution Hamiltonian in the reference frame of the specified conditions $t\mapsto X_\mu(t)$ and $\bar{f}>0$, for the ``classical ADM theory" governed by the constraints $\{\bar{C}_\mu\}$. In general, the system $\{\bar{C}_\mu\}$ would be deformed from the ADM constraints and do not form a first class system, mainly due to the quantum anomaly introduced to the highly nonlinear quantum constraints $\{\widehat{C}_\mu\}$. Nevertheless, whenever $\Omega_{cl}$ exists under a quantum reference frame our evolution Hamiltonian is of the same form as given in \eqref{intform2}-- only with the ${P}_{\mu,1}$ being the solution of the deformed constraints $\{\bar{C}_\mu\}$ rather than of the first-class ADM constraints.

The ``quantum corrections" in this series expansion lie in the terms ${\Delta}^{\scriptscriptstyle{k}}G^{\scriptscriptstyle{(N,r)}}_{\scriptscriptstyle{\!\widehat{H}_t}}$ with $(N,r)\neq (0,0)$ in \eqref{87}, given by the contributions to \eqref{7} with the terms involving ${\Delta}^{\!\scriptscriptstyle{k}}G^{\scriptscriptstyle{(N,r)}}_{\scriptscriptstyle{\widehat{ P}_{t_1,t_2}}}$ with $(N,r)\neq (0,0)$, which are directly computable using \eqref{semi1} and \eqref{33}. For the region outside of ${\Omega_{scl}}$ where we have ${}^{\scriptscriptstyle{\infty}}G^{\scriptscriptstyle{(0,0)}}_{\scriptscriptstyle{{\widehat{ P}_{t_2,t_1}}}}=0$, these quantum correction terms become the only contributors to $G_{\scriptscriptstyle{\!\widehat{H}_t}}$.

\section{Discussion}

\subsection*{Role of the anomalous quantum constraint system $\{\widehat{C}_\mu\}$ in the form of $G_{\scriptscriptstyle{\!\widehat{H}_t}}$}

One of the most contested subjects in Dirac quantum gravity is the prevailing quantum anomaly \cite{19.1,15.2} in the commutator algebra of the quantum constraints $\{\widehat{C}_\mu\}$ mentioned above. To tackle the quantization of the highly non-linear ADM constraints, novel ideas have been introduced to constructing the quantum constraint operators $\{\widehat{C}_\mu\}$ rigorously, for example via the flux-holonomy formulation in loop quantum gravity \cite{8.2 lqg,15.3}. These breakthroughs have led to new insights on the microscopic nature of gravity as described by the Hilbert space supporting the new quantum constraints, such as the LQG spin-network states describing the spatial quantum geometry. On the other hand, for the well-definedness the new quantum constraints seem inevitably come with the anomalous deformations leaving their commutator algebra often not even closed.

The first-class property of the ADM constraint system in canonical GR guarantees that the constraints arises from the fundamental symmetry of the theory \cite{1.1,1.2}, so that the physical degrees of freedom can be captured by the so-called ``reduced phase space”. Due to the first-class property, the ADM constraints generate the constraint-preserving symmetry transformations, over which the ``constraint surface" in the phase space can modularized into the ``reduced phase space" of the physical degrees of freedom for the evolution. Moreover, the form of the ADM constraint Poisson algebra also ensures the further ``foliation consistency" of the Hamiltonian dynamics in the reduced phase space. For a properly chosen reference frame with $t\mapsto X_\mu(t)$ and in its valid region with $f_{cl}(X_\mu,P_\mu)>0$, the system $\{{C}_\mu\}$ may be re-written as the equivalent constraint system $\{{P}_{\mu}-\mathcal{P}_{\mu,1}(X_I,P_I,X_\mu)\}$, with the functions $\mathcal{P}_{\mu,1}$ clearly being the solutions of ${P}_{\mu}$ in the constraint surface. Suppose the $\mathcal{N}_\mu(t){P}_{\mu}$ with a given $t\mapsto \mathcal{N}_\mu(t)$ generates the trajectory of $t\mapsto X_\mu(t)$, the corresponding evolution Hamiltonian then emerges from the ADM constraints in the form $\mathcal{H}_t(X_I,P_I)\equiv\mathcal{N}_\mu(t)\,\mathcal{P}_{\mu,1}(X_I,P_I,X_\mu(t))$. Now, if there are two reference frames with their frame conditions coinciding only at the initial and the final moments, the overall evolutions from the initial to final moments by their respective evolution Hamiltonians must be identical. This consistency-- protected by the ADM constraint Poisson algebra-- then encodes the notion of the frame-independent spacetime, for which the two evolutions are simply the descriptions of the same spacetime under the two spatial Cauchy-surface foliations coinciding in the initial and final slices.

If the physical degrees of freedom and their unitary evolutions in the Dirac quantum gravity are to be defined by the commutator flows of an equivalent quantum constraint system to $\{\widehat{C}_\mu\}$ in a directly analogous manner, the anomalous deformations in the constraint algebra would fundamentally threaten the desired Schr\"odinger dynamics capable of recovering the reduced phase space evolutions of the frame-independent spacetime. In our context, the problems of the anomaly can be concretely studied through the quantum reference frame approach, for which the degrees of freedom and the evolutions are defined and generated by instead the relevant transition maps $\widehat{\mathbb{P}}_{tt'}$. 

For any proposed assignment of $t\to \mathbb{K}_t$, our construction implies that the above quests may be achieved if the maps $\widehat{\mathbb{P}}_{tt}$ and $\widehat{\mathbb{P}}_{t,\scriptscriptstyle{[t,t']}}$ described in \eqref{uni cond} are both injective, just so that the assignment gives a valid quantum reference frame. Indeed, from the isomorphism condition we see that the described subspace $\mathbb{H}_t\subset \mathbb{H}$ is isometric to the $\mathbb{K}_t$ representing the corresponding classical reduced phase space with the coordinates of $(X_I,P_I)$. The issue of the foliation consistency is further addressed by the domain stability condition; here, the unitary evolution is given by the map $\widehat{U}_{t_2t_1}$ bridging the $\mathbb{K}_{t_1}$ and $\mathbb{K}_{t_2}$ as the two isometric representations of $\mathbb{D}_{t_1}$.  Our unitary evolution is thus by construction ``foliation consistent" with the $\mathbb{K}_{t_1}$ playing the role of a ``quantum Cauchy surface"; it is easy to see that if two quantum reference frames $t\to \mathbb{K}_t$ and $\tau\to \mathbb{K}_\tau$ coincides at the surfaces $(\mathbb{K}_{t_1}, \mathbb{K}_{t_2})=(\mathbb{K}_{\tau_1}, \mathbb{K}_{\tau_2})$ then we have the identical overall evolutions $\widehat{U}_{t_2t_1}= \widehat{U}_{\tau_2 \tau_1}$ for any state $\Psi\in \mathbb{H}_{t_1}$ as a frame independent quantum spacetime. 

Lastly, although the evolution Hamiltonian operator is not obtained through an equivalent set of the quantum constraints, we have shown that they are approximately so in terms of the Weyl symbols for the semi classical limits. In view of our result \eqref{intform2}, the lowest-order symbol ${}^{\scriptscriptstyle\infty}G^{\scriptscriptstyle{(0,0)}}_{\scriptscriptstyle{\!\widehat{H}_t}}$ differs from its classical counterpart just by the deviations of the solutions ${P}_{\mu,1}$ from the $\mathcal{P}_{\mu,1}$, due to the deformations of the set $\{\bar{C}_\mu\}$ from the original ADM constraints.

\subsection*{First principle dynamics in quantum cosmology using $G_{\scriptscriptstyle{\!\widehat{H}_t}}$}

One important application of our treatment is for the quantum cosmology in the deep quantum or strong gravity regimes, when the fullest interactions must be considered. Such models face the same challenges of obtaining the Schrodinger evolution without a fixed spacetime background. For example, describing the inhomogeneous perturbations of the universe upon the quantum geometry of the FRW LQC, the inflationary cosmic perturbation models of LQC \cite{17.1,17.2} have suggested intriguing corrections to the primordial perturbation power spectrum (such as the enhanced non-Gaussian statistics for certain modes) due to the discretized quantum geometry. As mentioned, these results of Schr\"odinger evolutions are mostly obtained by ignoring the perturbation’s back-reactions to the unperturbed spacetime and truncating certain quantum fluctuations. After such simplifications the original quantum constraints of the model can be factorized to give the Schrödinger equation for the perturbations in the FRW spacetime serving as a fixed background. 

More recently the non-linear order perturbation effects in these models are being studied, for the signals in the large scale perturbation spectrum and the Non-Gaussianity \cite{17.2}. However, to thse higher order perturbations, the back-reactions and the complete quantum gravitational fluctuations can become important, thus the challenge here is to derive the Schrödinger dynamics directly from the original quantum constraints encoding the fullest interactions, which can be then consistently truncated to arbitrary order of perturbation. Our approach offers such a natural solution. The existing models are mostly based on the elementary set of $$\{\widehat{X}_i, \widehat{P}_i\}\equiv \{\widehat{X}_a, \widehat{P}_a\} \cup \{\widehat{\delta X}_b,\widehat{\delta P}_b\},$$ where the subset $\{\widehat{X}_a, \widehat{P}_a\}$ are the $U(1)$ flux-holonomy operators on a single-vertex graph describing the spatial quantum geometry of FRW LQC, which is interacting with the inhomogeneous perturbation modes described by the canonically quantized $\{\widehat{\delta X}_b,\widehat{\delta P}_b\}$. Just like the full theory, such a model would be defined with the original quantum constraints $\{\widehat{C}_\mu(\widehat{X}_i, \widehat{P}_i)\}$, from which we may compute the evolution Hamiltonian in each quantum reference frame. Crucially, the truncation to any perturbation order can always be done as a final step upon the series expansion of the evolution Hamiltonian. Indeed in this context with $(\widehat{\delta X}_b,\widehat{\delta P}_b)\equiv(\gamma\widehat{\delta X}_b,\gamma\widehat{\delta P}_b)_{\gamma=1}$, the expansion \eqref{semi1} would be the series over $\lambda^r\gamma^p$, and correspondingly ``first-principle" evolution Hamiltonian in a quantum reference frame would be obtained in the series expansion of the form 
\begin{eqnarray}
G_{\scriptscriptstyle{\!\widehat{H}_t}}(X_I,\delta X_J, P_I, \delta P_J)
\equiv \sum_{k,N,r,p=1}^\infty \eta^{N}\lambda^r\gamma^p\, \,{\Delta}^{k}G^{\scriptscriptstyle{(N,r,p)}}_{\scriptscriptstyle{\!\widehat{H}_t}} |_{\eta,\lambda,\gamma=1}\nonumber 
\end{eqnarray}
for which the truncation to an arbitrary perturbation order of $p$ truly contains the full interactions (including the back-reactions) in the original constraints to that order. This treatment may thus be of great values in testing the first-principle quantum geometric effects via the higher-order quantum cosmic perturbations.

Another important subject is the dynamics of the black hole formation via the gravitational collapse. Describing the spherically symmetric spacetime with the spatial quantum geometry, the models of spherically symmetric LQC \cite{17.3,17.4} have suggested certain signature predictions to the black hole formation (such as the black-to-white-hole singularity resolutions). In comparison to the cosmic perturbation models, here the relation is even more obscured between the original quantum constraints and the desired Schrödinger propagators. Due to the difficulty, the dynamics is mostly studied either using the “effective classical treatment” with the “effective ADM constraints” given by the expectation values of the quantum constraints, or using a restricted model describing only the black hole interior where the quantum constraints may easily simplify to the Schrödinger form. 

To study the dramatic evolutions of the gravitational collapse, one ultimately should not assume the pre-existing horizon interior or the smallness of the quantum gravitational fluctuations. Thus the challenge here is again to derive the Schrödinger dynamics directly from the original quantum constraints encoding the fullest interactions. The general kinematics of the spherically symmetric loop quantum gravity has been constructed with the elementary set $\{\widehat{X}_i, \widehat{P}_i\}$ being the $U(1)$ flux-holonomy operators defined on a one-dimensional graph along the ``radius direction", describing the spherically symmetric spatial quantum geometry and the appropriate matter fields. The models are given the original quantum constraint system $\{\widehat{C}_\mu(\widehat{X}_i, \widehat{P}_i)\}$, with the mentioned obstacles due to their intricate forms. The result in this work can be applied to such models to obtain the evolution Hamiltonian $G_{\scriptscriptstyle{\!\widehat{H}_t}}(X_I, P_I)$ for a quantum reference frame $t\mapsto \mathbb{K}_t$. Then, one may simply select a proper initial Schr\"odinger state $\Psi_{(0)}(X_I)$ sharply peaked at a value of  $(\bar{X}_I, \bar{P}_I)$ describing a slight localized concentration of matter in the almost flat spacetime region, and then generate its evolution numerically with the $G_{\scriptscriptstyle{\!\widehat{H}_t}}$. The outcome of this computation—hopefully showing the event horizon testable by the causal correlation in \eqref{n point1} – would be the genuine first-principle prediction of the original quantum constraints $\{\widehat{C}_\mu\}$ for the evolution of the complete quantum Cauchy surface.

\subsection*{ Wilsonian renormalization of ${G}_{\widehat{H}_t}$ via effective quantum reference frames}

The classical limit in the region ${\Omega_{cl.}}$ introduced in this work is meaningful for the physical states sharply peaked for the elementary relational operators $\{\widehat{X}_I(t), \widehat{P}_I(t)\}$ with small wave-function spreadings. For a fundamental Dirac theory with the elementary operators representing the Planck-scale variables, such limit thus gives the “effective classical theory" of the Planck-scale. Since our current universe is instead dominated by the microscopic quantum fluctuations, its effective classical laws in the large scales would be driven away from the Planck-scale classical limit by the quantum fluctuations. In our context, being able to compute the Planck-scale Schr\"odinger dynamics for a fundamental Dirac theory of QG, we need a renormalization method for obtaining the effective theory of our universe at its current scale of quantum fluctuations \cite{19,20,21}. The following is our proposal for the next step.

Suppose we are given the Planck-scale Schrodinger theory in a quantum reference frame. Assume that its elementary operators can be devided into two commuting subsets $$\{\widehat{X}_I,\widehat{P}_I\}\equiv \{\widehat{X}^{\rho}_{a},\widehat{P}^{\rho}_{a}\} \cup\{ \widehat{X}^{\rho}_{J},\widehat{P}^{\rho}_{J} \}$$ where the first subset is regarded as the microscopic degrees of freedom at the scale $\rho$. In the canonical quantum field theories, the set $\{\widehat{X}^{\rho}_{a},\widehat{P}^{\rho}_{a}\}$ may represent all the field Fourier modes with the wave numbers larger than $\rho$. 

Let us observe that our construction naturally incorporate the existing Hamiltonian renormalization methods \cite{22,23} using the same language of Hilbert space isometric representations. To see this, given a certain $\rho$ let us introduce the ``effective Cauchy surface" $$\mathbb{S}^{\scriptscriptstyle{\rho}}_t \equiv span\{ \ket{T_{\mu}(t),\phi^\rho}\otimes\ket{ {X}^{\rho}_{J}}\}\subset \mathbb{S}_t$$  with a properly chosen disentangled microscopic wave function $\phi^\rho({X}^{\rho}_{a})$. If we were describing a scattering-particle system with the energies of the in-coming particles bounded by $\rho$, the state $\phi^\rho$ may be given by the Fock vacuum of the ``bare particles" with energies way-beyond $\rho$. Let us assume that each effective state in $\mathbb{S}^{\scriptscriptstyle{\rho}}_t$ can be ``dressed” into its associate physical Schr\"odinger state by the unitary "dressing map" $$(\widehat{\beta}_t^{\scriptscriptstyle \rho\,\dagger})^{-1}=\widehat{\beta}^{\scriptscriptstyle \rho}_t: \mathbb{S}_t \to \mathbb{S}_t$$ which defines the associated spaces of $\mathbb{S}^{\scriptscriptstyle{\rho}}_t\equiv Image(\widehat{\beta}^{\scriptscriptstyle \rho}_t \restriction_{\mathbb{S}^{\scriptscriptstyle{\rho}}_t})$ and $\mathbb{D}^{\scriptscriptstyle{\rho}}_t\equiv Image(\widehat{\mathbb{P}} \restriction_{\mathbb{S}^{\scriptscriptstyle{\rho}}_t})\subset \mathbb{D}_t$. Now, suppose for a fixed $\phi^\rho$ there is a special dressing map $t\mapsto\widehat{\beta}^{\scriptscriptstyle \rho}_t$ for each moment $t$, such that the effective domain stability holds in the form
\begin{eqnarray}
\label{effuni}
\mathbb{D}^{\scriptscriptstyle{\rho}}_t= \mathbb{D}^{\scriptscriptstyle{\rho}}_{t_1}\, ; \, t\in [t_1,t_2].
\end{eqnarray}
We may then compute the fundamental Schr\"odinger evolution of any physical state $\Psi^{\scriptscriptstyle{\rho}} \in \mathbb{D}^{\scriptscriptstyle{\rho}}_{t_1}$ using the effective theory, by replacing the fundamental isometry $\widehat{I}_t$ by the effective isometry $\widehat{I}^{\scriptscriptstyle{\rho}}_t\equiv \widehat{I}_t\,\widehat{\beta}^{\scriptscriptstyle \rho}_t$ throughout \eqref{isom} to \eqref{wave func} for the effective elementary relational operators $\{\widehat{X}^{\rho}_{J}(t),\widehat{P}^{\rho}_{J}(t) \} \equiv\{ \widehat{\beta}^{\scriptscriptstyle \rho}_t\widehat{X}_{J}\widehat{\beta}^{\scriptscriptstyle \rho \,-1}_t(t)\,,\, \widehat{\beta}^{\scriptscriptstyle \rho}_t\widehat{P}_{J}\widehat{\beta}^{\scriptscriptstyle \rho \,-1}_t(t)\}$. 

Under the effective quantum reference frame thus defined as $$t\mapsto ( \mathbb{S}^{\scriptscriptstyle{\text{eff} \,\rho}}_t\subset \mathbb{S}_t\,, \,\widehat{\beta}^{\scriptscriptstyle \rho}_t)$$ satisfying \eqref{effuni}, the physical state $\Psi^{\scriptscriptstyle{\rho}}$ can be described by the Schr\"odinger wave function of the form $\Psi_{\scriptscriptstyle{(t)}}^{\scriptscriptstyle{\rho}}({\mathcal{X}}^{\rho}_{I})$, with $\widehat{\mathcal{X}}^{\rho}=\widehat{\mathcal{X}}^{\rho\,\dagger}=\widehat{\mathcal{X}}^{\rho}(\widehat{X}^{\rho}_{J}, \widehat{P}^{\rho}_{J})$, evolving unitarily during $[t_1,t_2]$ by the effective propagator 
\begin{eqnarray}
\label{propeff}
G_{\widehat{U}^{\scriptscriptstyle \rho}_{t',\,t}}
= tr_{\scriptscriptstyle \rho}\big[\,G_{\ket{\phi^\rho}\bra{\phi^\rho}} \,\,G^*_{\scriptscriptstyle{\widehat{\beta}^{{\scriptscriptstyle \rho}}_{t'}}}\star G_{\widehat{U}_{ t't}}\star G_{\scriptscriptstyle{\widehat{\beta}^{{\scriptscriptstyle \rho}}_t}}\,\big]\,,
\end{eqnarray}
where the partial trace $tr_{\scriptscriptstyle \rho}$ with the effectively decoupled $\phi^\rho$ is taken over the microscopic sector, and it clearly relates to the ``integrating out" for the microscopic degrees of freedom in the familiar functional renormalization. Here the effective evolution Hamiltonian $ G_{\widehat{H}^{\scriptscriptstyle \rho}_{t}}({X}^{\rho}_{I},{P}^{\rho}_{J})$ would be then given by 
\begin{eqnarray}
\label{propeff2}
G_{\widehat{H}^{\scriptscriptstyle \rho}_{t}}=-i\hbar(\partial_{t_2}-\partial_{t_1}) G_{\widehat{U}^{\scriptscriptstyle \rho}_{t_2,t_1}}\restriction_{t_1=t_2=t}.
\end{eqnarray}
The task of searching for the effective theory at the scale $\rho$ is now formulated as searching for the desired map of $$(\rho,t)\mapsto (\,G_{\ket{\phi^\rho}\bra{\phi^\rho}}\,, \,G_{\scriptscriptstyle{\widehat{\beta}^{{\scriptscriptstyle \rho}}_t}}\,) $$. This map should satisfy the definition conditions of 
\begin{eqnarray}
\label{renorm eq}
G_{\scriptscriptstyle{\widehat{\beta}^{\rho_{Planck}}_t}} =1 \,,\,G^*_{\scriptscriptstyle{\widehat{\beta}^{{\scriptscriptstyle \rho}}_t}}\star G_{\scriptscriptstyle{\widehat{\beta}^{{\scriptscriptstyle \rho}}_t}}=1\, ,\, ...\,
\end{eqnarray}
and also the dynamic conditions of
\begin{eqnarray}
\label{renorm eq2}
\partial_\rho\,\,\, Tr\big[\,G_{\ket{\psi_{(\rho)}}\bra{\psi'_{(\rho)}}} \,\,(G_{\widehat{H}^{\scriptscriptstyle \rho}_{t}}-G^*_{\widehat{H}^{\scriptscriptstyle \rho}_{t}})\big]=0\,,\,...\,,
\end{eqnarray}
containing the effective domain stability \eqref{effuni} manifesting as the self adjointness of $\widehat{H}^{\scriptscriptstyle \rho}_{t}$, where the trace is taken with the arbitrarily assigned probing wave-functions $\rho \mapsto (\psi_{(\rho)}({X}^{\rho}_{J}), \psi'_{(\rho)}({X}^{\rho}_{J}))$. Viewed with \eqref{propeff} and \eqref{propeff2}, the complete conditions above could lead to a ``renormalization flow equation" in the phase space functional form of ($r,r'=0,1$)
\begin{eqnarray}
\label{ren}
\mathcal{F}\big[\,\partial^r_t\partial ^{r'}_\rho G_{\scriptscriptstyle{\widehat{\beta}^{{\scriptscriptstyle \rho}}_t}}\,, \,G_{\ket{\phi^\rho}\bra{\phi^\rho}}\,, \,G_{\widehat{H}^{\scriptscriptstyle \rho}_{t}}\,\big]=0
\end{eqnarray}
which allows us to propagate $G_{\scriptscriptstyle{\widehat{\beta}^{{\scriptscriptstyle \rho}}_t}}$ to any scale in $\rho$ from its given Planck-scale value of unity, using the given $\rho\mapsto G_{\ket{\phi^\rho}\bra{\phi^\rho}}$ and $G_{\widehat{H}_{t}}$. Through supplementing the physical conditions \eqref{renorm eq} and \eqref{renorm eq2} for the renormalization flow equation, we expect our formulation to be instrumental in bridging the existing Hamiltonian renormalization approach to that for the fundamental Dirac quantum gravity. Also, due to the explicit agreement \cite{13} between our Schr\"odinger propagator and the corresponding canonical Fadeev-Popov path integral, our method of renormalization may also be studied in direct contact with the functional renormalization methods.

\section{Summary and conclusion}

For a Dirac theory of quantum gravity governed by the quantum master constraint $\widehat{M}$ via the the rigging map $\widehat{\mathbb{P}}=\delta(\widehat{M})$, we apply our approach of quantum reference frame in the Wigner-Weyl representation and successfully obtained the explicit expression of the evolution Hamiltonian operator ${G}_{\widehat{H}_t}$ for the large class of the quantum reference frames, directly in terms of the quantum constraint $G_{\scriptscriptstyle{\widehat{M}}}$ and the reference-frame conditions $\{G_{\scriptscriptstyle{\widehat{f}}}\,, \,G_{ \scriptscriptstyle{\ket{{X}_\mu(t)}\bra{{X}_\mu(t)}} }\}$.

As stated, the construction still needs a more concrete guidence and physical interpretation for identifying the valid quantum reference frames. Also, beyond the preliminary series expansion scheme, it remains to be studied for the best expansion scheme in the context of a specific model or theory, for the optimized series convergence and the physical meaning of the expansion. Moreover, in the application to the full theory of the non-canonical cases we may need to first formulate the Wigner-Weyl representation for its kinematic Hilbert space, such as to extend the flux-holonomy Wigner-Weyl representation to the $SU(2)$ setting. 

Nevertheless, our result may be a significant step toward the goal described in the begining: a fundamental approach for the unitary evolution of the genuine observables in the Dirac theory of quantum gravity, exactly computable with the only inputs being the governing quantum master constraints and the quantum reference frame conditions. Also, importantly, such direct computation allows one to obtain the full dynamics of a physical state in the language of the relational elementary operators for the frame, while implicitly bypassing the (almost impossible) difficulties in finding the solution space of the quantum constraints.

Many important yet elusive subjects in QG could now be studied with calculations at the first-principle level, and several of them are discussed in the last section. Our results may lead to a new and concrete way to explore the first-principle dynamical content of the quantum constraint system $\{\widehat{C}_\mu\}$.

\section*{Dedication}
With love and gratitude, the author warmly dedicates this work to his parents.

\begin{acknowledgments}
The author wish to thank Professor Yeo-Yie Charng and Professor Steven Carlip for their helpful discussions and suggestions.
\end{acknowledgments}

\appendix

\section{Diamond expansion for the $\star$-polynomials}

Here we demonstrate the relation \eqref{000} to be true. Due to the multi-linearity of the $\star$-products and \eqref{0011}, it suffices to prove the relation for the special cases with $\{\widehat{f}_n= (\widehat{A}_n)^{m_n}\}_{n=1,2,..,N}$. Here we want to compute $$ G_{\scriptscriptstyle{\widehat{f}_1}}\star G_{\scriptscriptstyle{\widehat{f}_2}}...\star G_{\scriptscriptstyle{\widehat{f}_N}}= \big[...\big[[1(\star\, G_{\scriptscriptstyle{\widehat{A}_1}})]^{m_1}(\star \,G_{\scriptscriptstyle{\widehat{A}_1}})]^{m_2}...(\star\, G_{\scriptscriptstyle{\widehat{A}_N}})\big]^{m_N}.$$ Using the decomposition $\star= \cdot\,+\, *$, we may organize the contributions to the power product by tracing the $*$- operations; note that since the $*$-operation itself is not associative, we have specified the decomposed product to be taken starting from the very left. 

For any form of the $\star$-product, we may always organized the decomposed contributions to the  above, by first summing over all the contributions with the specified numbers $\{\tilde{m}_n\leq m_n\}$ of the individual factors of $G_{\scriptscriptstyle{\widehat{A}_n}}$ acted upon by at least one of the $*$-operations.

The $\{\tilde{m}_n=0\}$ term in this expansion is thus given by the single contribution in the decomposition, in which every individual factor of $G_{\scriptscriptstyle{\widehat{A}_n}}$ of the polynomial is acted upon only by the $\cdot\,$-product (the c-number product), which just gives $(G_{\scriptscriptstyle{\widehat{A}_1}})^{m_1}(G_{\scriptscriptstyle{\widehat{A}_2}})^{m_2}...(G_{\scriptscriptstyle{\widehat{A}_N}})^{m_N}$.

Next, we look into the decomposed contributions of a given $\{\tilde{m}_n\}$ with non zeroes. Each of such contribution is simply associated to a distinct selection of $\tilde{m}_n$ members from the $m_n$ individual factors of $G_{\scriptscriptstyle{\widehat{A}_n}}$, to be acted upon by at least one of the $*$-operations. Observe that, since the unselected $({m_n-\tilde{m}_n})$ members of the individual  $G_{\scriptscriptstyle{\widehat{A}_n}}$ factors are acted upon by only the $\cdot\,$-product, all the contributions for the specified $\{\tilde{m}_n\}$ would have the same c-number prefactor given by $ \prod_{n=1}^N (G_{\scriptscriptstyle{\widehat{A}_n}})^{m_n-\tilde{m}_n}$. Following the common prefactor the remaining parts are also the same: given by first taking the decomposed $\cdot\,+\, *$ product between the selected factors in the original sequence, and removing all the terms with any selected factor acted upon only by the $\cdot$ product. This is just the contraction defined in \eqref{ctr} between the selected factors. Finally, since the multiplicity of decomposed contributions for a certain $\{\tilde{m}_n\}$ is just the counting of the distinct selections the $\tilde{m}_n$ individual factors, given by ${{m_n}\choose{\tilde{m}_n}}$, we conclude that all the decomposed contributions with a certain $\{\tilde{m}_n\}$ must sum up to
\begin{eqnarray}  
{{m_n}\choose{\tilde{m}_n}}\prod_{n=1}^N (G_{\scriptscriptstyle{\widehat{A}_n}})^{\scriptscriptstyle{m_n-\tilde{m}_n}}\, \{(G_{\scriptscriptstyle{\widehat{A}_1}})^{\tilde{m}_1}(G_{\scriptscriptstyle{\widehat{A}_1}})^{\tilde{m}_2}...(G_{\scriptscriptstyle{\widehat{A}_N}})^{\tilde{m}_N}\}_{ctr}\nonumber\\
=\prod_{n=1}^N  \partial^{\tilde{m}_n}_{G_{\scriptscriptstyle{\widehat{A}_n}}} f_n(G_{\scriptscriptstyle{\widehat{A}_n}})\frac{\{(G_{\scriptscriptstyle{\widehat{A}_1}})^{\tilde{m}_1}(G_{\scriptscriptstyle{\widehat{A}_1}})^{\tilde{m}_2}...(G_{\scriptscriptstyle{\widehat{A}_N}})^{\tilde{m}_N}\}_{ctr}}{\tilde{m}_1!\tilde{m}_2!\,...\,\tilde{m}_N!} .\nonumber\\
\end{eqnarray}
Further summing over the above with all the possible values in $\{\tilde{m}_n\}$, we obtain the full diamond expansion for the case $\{\widehat{f}_n= (\widehat{A}_n)^{m_n}\}_{n=1,2,..,N}$ as 
\begin{eqnarray}  
G_{\scriptscriptstyle{\widehat{f}_1}}\star G_{\scriptscriptstyle{\widehat{f}_1}}...\star G_{\scriptscriptstyle{\widehat{f}_N}}=G_{\scriptscriptstyle{\widehat{f}_1}}\diamond G_{\scriptscriptstyle{\widehat{f}_1}}...\diamond G_{\scriptscriptstyle{\widehat{f}_N}}.
\end{eqnarray}
By the linearity this result also extends to the cases of \eqref{0011}.

\end{document}